\documentclass[pre,twocolumn,9pt]{revtex4}

\newcommand{\dd}{\mathrm{d}}
\newcommand{\ii}{\mathrm{i}}

\newcommand{\vettoreErre}{\boldsymbol{r}}

\newcommand{\CalA}{\mathcal{A}}

\usepackage{ae} 
\usepackage{amsmath}
\usepackage{amsfonts}
\usepackage{hyperref}
\usepackage{graphicx}
\usepackage[T1]{fontenc}

\begin{document}

\title{Paraxial Sharp-Edge Diffraction of Vortex Beams by Elliptic Apertures}

\author{Riccardo Borghi}
\affiliation{Dipartimento di Ingegneria Civile, Informatica e delle Tecnologie Aeronautiche, \\
Universit\`{a} ``Roma Tre'', Via Vito Volterra 62, I-00146 Rome, Italy}


\begin{abstract}
A semi-analytical computational algorithm to model the wavefield generated by  paraxial diffraction 
of a class of Laguerre-Gauss beams by sharp-edge elliptic apertures is here developed. 
{ Thanks to such a powerful computational tool,} some basic aspects of an intriguing 
and still unexplored  singular optics scenario can be studied, within a geometry as simple as possible, 
with arbitrarily high accuracies.
\end{abstract}

\maketitle

Light beams carrying out topological singularities in the form of phase vortices continue to be among the most investigated 
``optical objects,'' since the pioneering work by Allen {\em et al.}~\cite{Allen/Beijersbergen/Spreeuw/Woerdman/1992,Soskin/Vasnetsov/2001,Dennis/OHolleran/Padgett/2009}.
The use of sharp-edge diffraction to unveil the hidden structure of phase singularities carried 
out by such vortex beams has inspired a great deal of work, both theoretical and experimental~\cite{Hickmann/Fonseca/Silva/ChavezCerda/2010,Bekshaev/Chernykh/Khoroshun/Mikhaylovskaya/2017,Chernykh/Petrov/2021}.
Most of the above papers explored a similar scenario, in which Laguerre-Gauss (LG) vortex beams were diffracted by hard-edge half planes. 
Sharp-edge diffraction of LG vortex beams has also been investigated with polygonal as well as circular apertures~\cite{Mourka/Baumgartl/Shanor/Dholakia/Wright/2011,Liu/Sun/Pu/Lu/2013,Ambuj/Vyas/Singh/2014,Stahl/Gbur/2016,Taira/Zhang/2017,Goto/Tsujimura/Kubo/2019,Narag/Hermosa/2019,Taira/Kohmura/2019,Rocha/Amaral/Fonseca/Jesus-Silva/2019,Borghi/2022}, although most of such investigations have been limited to far-field (Fraunhofer) analysis.

A different  scenario, which seems to be still largely unexplored so far, has to do with sharp-edge diffraction of vortex beams by {\em  noncircular} smooth apertures. 
To point out its relevance, a few words by Michael Berry~\cite{Berry/2001} deserve to be quoted:
``for general smooth boundaries, focusing of edge waves occurs on caustic curves (envelope of normals to the boundary).''  
Berry's words must be contextualized within the framework of the so-called Catastrophe Optics~\cite{Berry/Upstill/1980,Nye/1999}.
Caustics generated from  plane-wave diffraction by smooth sharp-edge apertures are approximately arranged along the geometric evolute of the aperture rim~\cite{Borghi/2015,Borghi/2016}. 
The possibility of using elliptic apertures to explore paraxial vortex sharp-edge diffraction represents the main idea of the present letter.

Ellipses have smooth, curved rims whose shapes are  controlled by a single real parameter (the eccentricity  or,  equivalently, the aspect ratio).
Ellipse's geometrical evolute is a symmetric closed curve called {\em astroid}, containing as many as four cusps. 
Ellipses are the simplest smooth closed curves  not endowed with  radial symmetry (except for a null eccentricity).  
They are natural candidates, together with LG beams, for building up a scenario mathematically as simple as possible, in which the interaction 
between singularities of different nature (optical vortices and caustics) can be explored.

In the first part of the  Letter, a semi-analytical algorithm will be developed to evaluate, up to arbitrarily high (in principle) accuracies, the optical field generated by the 
paraxial diffraction of a monochromatic (wavelength $\lambda$) vortex LG beams carrying out a given topological charge $m\in\mathbb{Z}$ by a sharp-edge elliptic aperture $\CalA$  of given aspect ratio $\chi\in (0,1)$,
placed in  a planar, opaque screen and delimited by the boundary $\Gamma=\partial \CalA$. 
The screen coincides with the plane $z=0$ of a suitable cylindrical reference frame $(\vettoreErre;z)$, whose $z$-axis coincides with the ellipse symmetry axis,
as well as with the mean propagation axis of the impinging LG beam. 
On denoting $\psi_0(\vettoreErre)$ the disturbance distribution of the latter at $z=0^-$, the field distribution, say $\psi$, at the observation point $P\equiv (\vettoreErre;z>0)$, 
is given, within paraxial approximation and apart from an overall phase factor $\exp(\ii k z)$, by 
the following dimensionless version of Fresnel's integral:
\begin{equation}
\label{Eq:FresnelIntegral}
\begin{array}{l}
\displaystyle
\psi\,=\,
-\frac{\mathrm{i}U}{2\pi}\,
\int_{\boldsymbol{\mathcal{A}}}\,
\mathrm{d}^2\,\rho\,
\psi_0(\boldsymbol{\rho})\,
\exp\left(\frac{\mathrm{i}U}{2}\,|\boldsymbol{r}-\boldsymbol{\rho}|^2\right)\,,
\end{array}
\end{equation}
where, in place of $z$, the Fresnel number $U=2\pi\,a^2/\lambda z$ has been introduced, the symbol $a$ denoting  a characteristic length of the aperture size,
which will be supposed to coincide with the ellipse major half-axis. All transverse lengths, both across the aperture and the observation 
planes, have been normalized to $a$: { this implies that ellipse's minor half-axis length equals $\chi$}. 
The  impinging LG beam distribution will  be modeled by the following function:
\begin{equation}
\label{Eq:Gaussian.1}
\begin{array}{l}
\displaystyle
\psi_0(\boldsymbol{\rho})\,=\,\rho^m\,\exp(\ii\,m\,\phi)\,\exp\left(\ii \dfrac\gamma 2 \rho^2\right)\,, 
\quad m \ge 0,\,\,\gamma \in \mathbb{C}\,,
\end{array}
\end{equation}
where $(\rho,\phi)$ denote (normalized) polar coordinates of the transverse vector $\boldsymbol{\rho}$ across the aperture plane,
while the overall amplitude factor has been set to the unity. 
The complex dimensionless parameter $\gamma$ gives account of the  curvature (via its real part), 
as well as the transverse spot size (via $\mathrm{Im}\{\gamma\} \ge 0 $) of the impinging beam. 
{ Moreover, the complex parameter $V=U+\gamma$, with $\mathrm{Re}\{\ii V\} \le 0$, will also be introduced.
When Eq.~(\ref{Eq:Gaussian.1}) is substituted into Eq.~(\ref{Eq:FresnelIntegral}), the resulting integral can 
be evaluated up to arbitrarily high (in principle) accuracies. Similarly as it was done in the past to deal with problems endowed with elliptic symmetry~\cite{{Kotlyar/Khonina/Almazov/Soifer/Jefimovs/Turunen/2006},Borghi/2014}, the key idea is to employ the following representation of Cartesian components of the transverse vectors $\boldsymbol{\rho}$ and $\boldsymbol{r}$:}
\begin{equation}
\label{Eq:EllipticHole.0.1}
\left\{
\begin{array}{l}
\displaystyle
\boldsymbol{\rho} \,=\, (\xi\,\cos\alpha\,,\,\chi\,\xi\,\sin\alpha)\,,\quad 0 \le \alpha \le 2\pi\,,\quad \xi \ge 0\,,\\
\boldsymbol{r} \,=\, (\chi\,X\cos\beta\,,\,X\sin\beta)\,,\quad 0 \le \beta \le 2\pi\,,\quad X \ge 0\,.
\end{array}
\right.
\end{equation}
%
{ Then, nontrivial algebra leads to the following representation of the diffracted wavefield $\psi$: }
\begin{equation}
\label{Eq:FresnelPropagatorConvolution.3}
\begin{array}{l}
\displaystyle
\psi\,=\,
\displaystyle
-\dfrac{\mathrm{i}U\chi}2\,\exp\left(\frac{\mathrm{i}}{2}U\,r^2\right)\,
\displaystyle\sum_{\ell=0}^m\,
\left({{m}\atop{\ell}}\right)\,\eta^\ell_+\eta^{m-l}_-\,\\
\\
\times
\Psi^m_{2\ell-m}\left[\dfrac{V(1+\chi^2)}{4},\,U\chi X,\,\dfrac{V(1-\chi^2)}4;\,\beta\right]\,,
\end{array}
\end{equation}
{ where $\eta_\pm=(1\pm\chi)/2$ and }
 the functions $\Psi^m_k$ are defined by
\begin{equation}
\label{Eq:FresnelPropagatorConvolution.3,1.1}
\begin{array}{l}
\displaystyle
\Psi^m_k(a,b,c;\varphi)\,=\,
\ii^{-k}\exp(\ii k\varphi)\,\sum_{n\in\mathbb{Z}}\,
\ii^{-n} \exp(\ii 2n\varphi)\\
\\
\times
\displaystyle
\int_0^1\,\mathrm{d}\xi\,\xi^{m/2}\,\exp(\ii a\xi)\,
J_n(c\xi)\,J_{2n+k}(b\sqrt\xi)\,.
\end{array}
\end{equation}
{ For room reason and in order to improve the readibility of the present Letter,
the mathematical proof of Eqs.~(\ref{Eq:FresnelPropagatorConvolution.3}) and~(\ref{Eq:FresnelPropagatorConvolution.3,1.1})  has been detailed in Appendix~\ref{Sec:IPart}.}

As we shall see in a moment, the numerical evaluation of the last integrals does not present any problems,
even for  nonsmall Fresnel's numbers and/or nonsmall values of $X$. 
Moreover,  as far as the convergence of the Fourier series into Eq.~(\ref{Eq:FresnelPropagatorConvolution.3,1.1}) is concerned, a truncation 
criterion similar to that proposed in Ref.~\cite{Borghi/2014} will be adopted, according to which the index $n$ will run within the range $[-N,N]$, where
\begin{equation}
\label{Eq:FourierApproach.6}
\begin{array}{l}
\displaystyle
N \,\simeq\, \mathrm{int}\left[\max\left\{U\chi X,\dfrac{|V|(1-\chi^2)}2\right\}\right]\,+\,1\,,
\end{array}
\end{equation}
with int$[\cdot]$ denoting the integer part operator. That is all.

In the second part of the Letter, the scenario above described will be initially explored with a LG beam with topological charge $m=3$. 
Such a value guarantees a reasonably high ratio between the values of the impinging optical intensity $|\psi_0|^2$ at the aperture boundary and those  in central region.
In this way, the main contribution to the diffracted beam $\psi$ is expected  to come from the rim $\Gamma$ (with the three singularities being initially all located at the beams axis).
{ For simplicity, the spot-size of the incident LG beams will be assumed  to be much greater of the aperture sizes, a choice which also represents a sort of ``computational worst case,'' 
as far as the evaluation of the integrals into Eq.~(\ref{Eq:FresnelPropagatorConvolution.3,1.1}) is concerned. In formulas, $\gamma\simeq 0$ and thus $V\simeq U$.}

Our exploration starts from the far-field limit $U\to 0$, 
where the diffracted field into Eq.~(\ref{Eq:FresnelIntegral}) takes on,
apart unessential overall phase and amplitude factors, the following  form:
\begin{equation}
\label{Eq:FourierIntegral}
\begin{array}{l}
\displaystyle
\psi_{\rm ff}\,=\,
\int_{{\mathcal{A}}}\,
\mathrm{d}^2\,\rho\,
\psi_0(\boldsymbol{\rho})\,
\exp\left(-\mathrm{i}\,U\,\boldsymbol{r}\cdot\boldsymbol{\rho}\right)\,,\qquad U \to 0\,.
\end{array}
\end{equation}
Equation~(\ref{Eq:FourierIntegral}) shows that $\psi_{\rm ff}$ depends only on the spatial frequency
$\boldsymbol{p}=U\boldsymbol{r}$. 
When $m=0$ (impinging plane wave), $\psi_{\rm ff}$ is proportional to the Fourier transform (FT) of the characteristic function of the elliptic aperture ${\mathcal{A}}$.
For $m>0$, the FT evaluation becomes considerably harder,  due to the mismatch 
between the different symmetries related to $\psi_0$ (radial) and to the aperture (elliptical, of course). 
{ Nevertheless, it is still possible to achieve exact, analytical expressions of the following Fourier integral:}
\begin{equation}
\label{Eq:FourierIntegral.1}
\begin{array}{l}
\displaystyle
\mathcal{F}_m(\boldsymbol{p})=\int_{\boldsymbol{\mathcal{A}}}\,
\mathrm{d}^2\,\rho\,
[\rho\,\exp(\mathrm{i} \phi)]^m
\exp\left(-\mathrm{i}\,\boldsymbol{p}\cdot\boldsymbol{\rho}\right)\,,\qquad m\in\mathbb{Z}\,.
\end{array}
\end{equation}
{ Again, the interested reader can be found the complete algorithm and the explicit result in Appendix~\ref{Sec:IPart}.}
Then, it is not difficult to prove that, in the far zone, the three singularities originally carried out by the impinging LG beam  
propagate along different rectilinear paths contained in the  $xz$-plane, with the $x$-axis being aligned along the ellipse 
major axis:  one of them (the so-called {\em on-axis} singularity) propagates along the $z$-axis. 
{ As far as the actual positions of the other two (the  {\em off-axis} singularities) are concerned, what should be expected 
is they would  approach, for sufficiently small values of $U$, the quantities $\left(\pm {\bar{p}_x}/U\,,\,0\right)$, with $\bar{p}_x$ being the least positive root of the equation $\mathcal{F}_3(\bar{p}_x,0)=0$. Such quantities will be referred to as the 
``far-field estimates''. }

{ In what follows, elliptic frames (in place of classical rectangular frames) will be employed to visualize amplitude and phase diffraction patterns. 
Such a technical expedient turns out to be useful to optimize the pattern visualization. 
In all figures, the elliptic frames will be chosen in order for all involved singularities, i.e., vortices and evolutes, to be included.}

\begin{figure}[!ht]
\centering
\begin{minipage}[t]{1.8cm}
\centerline{\includegraphics[width=1.8cm,angle=-0]{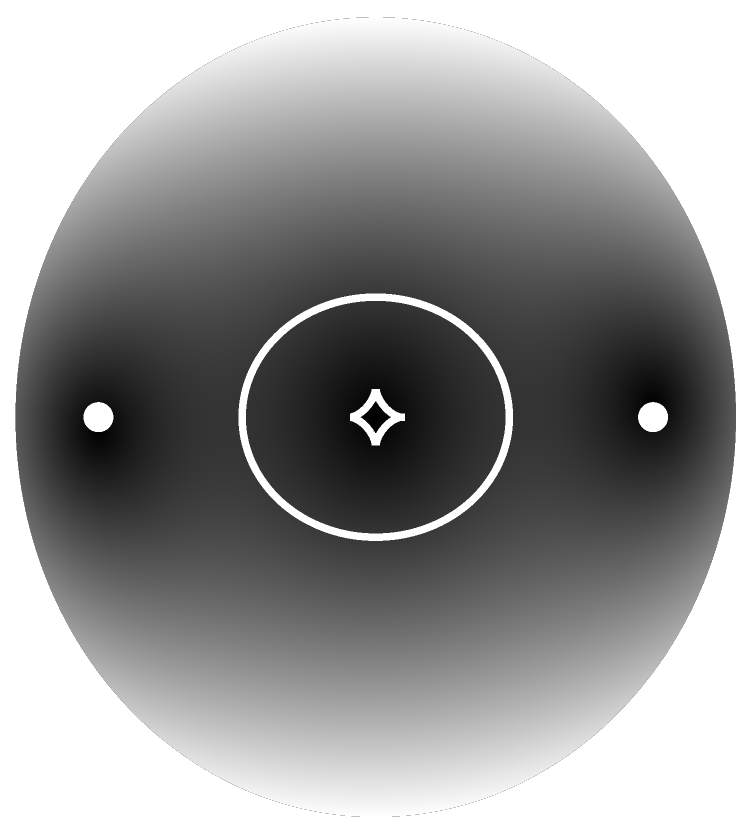}}
\centerline{\includegraphics[width=1.8cm,angle=-0]{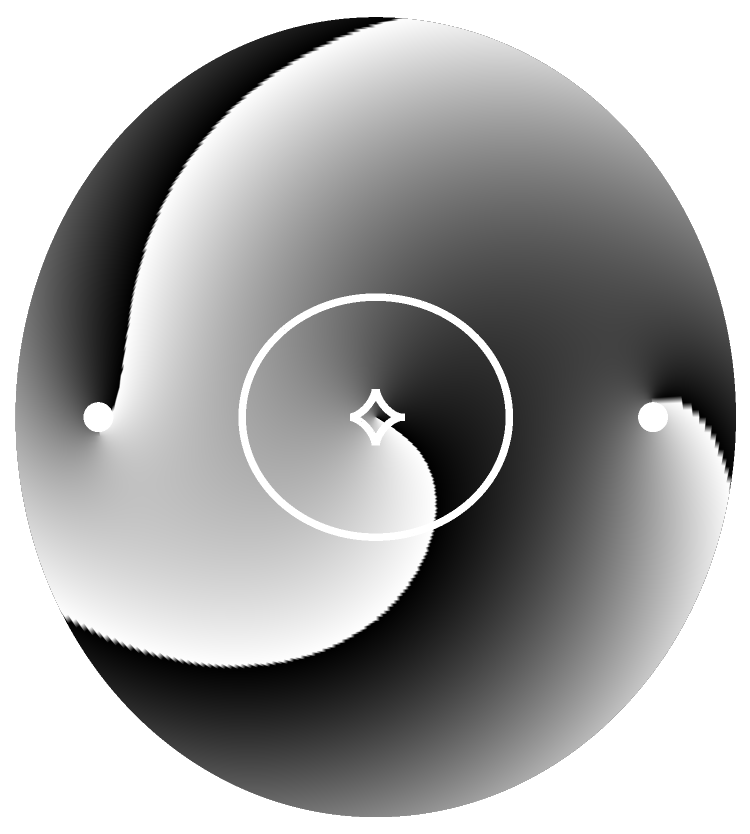}}
\end{minipage}
\hspace*{1mm}
\begin{minipage}[t]{1.8cm}
\centerline{\includegraphics[width=1.8cm,angle=-0]{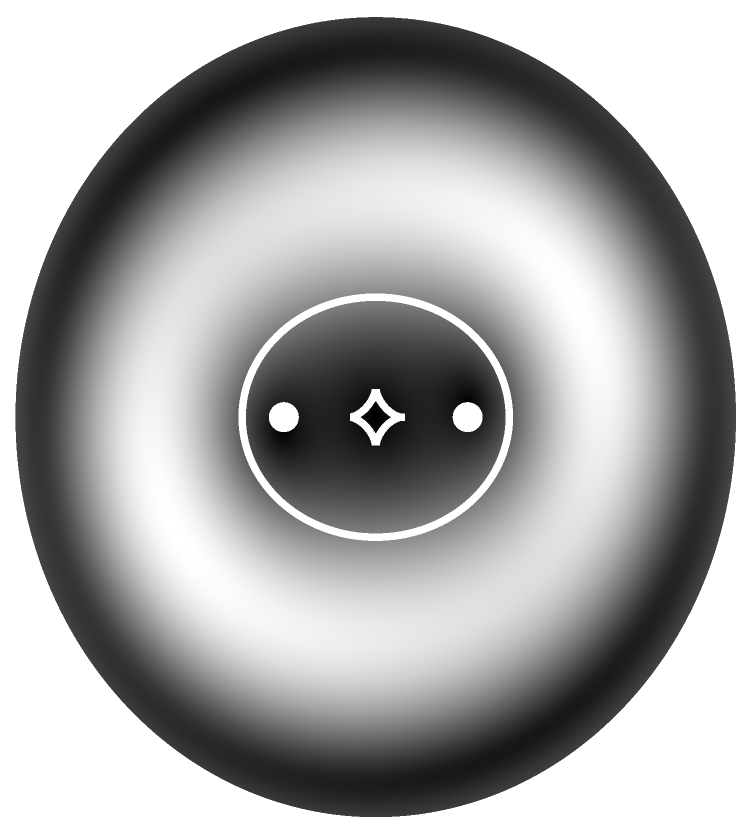}}
\centerline{\includegraphics[width=1.8cm,angle=-0]{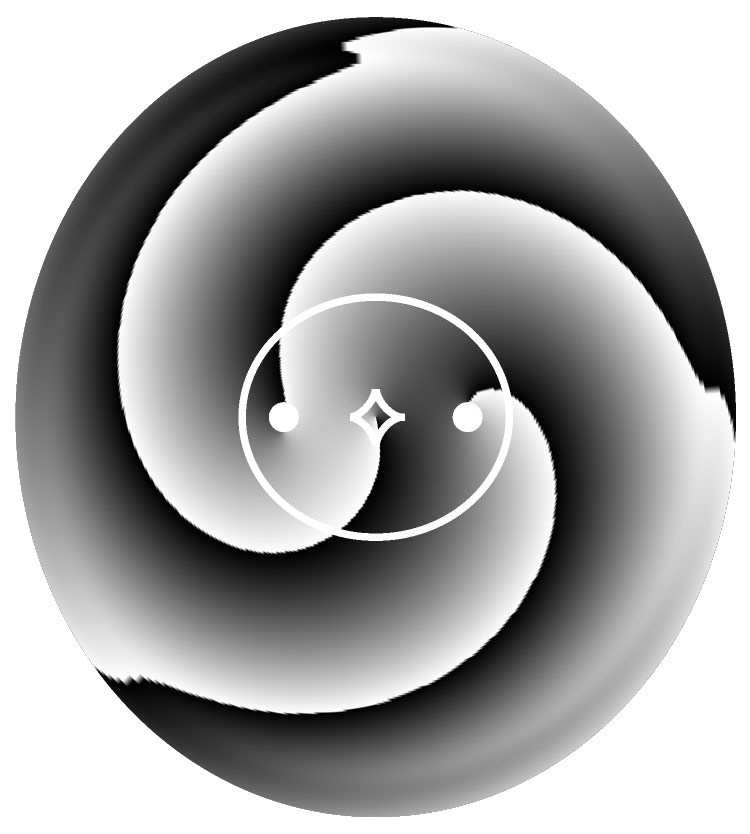}}
\end{minipage}
\hspace*{1mm}
\begin{minipage}[t]{1.8cm}
\centerline{\includegraphics[width=1.8cm,angle=-0]{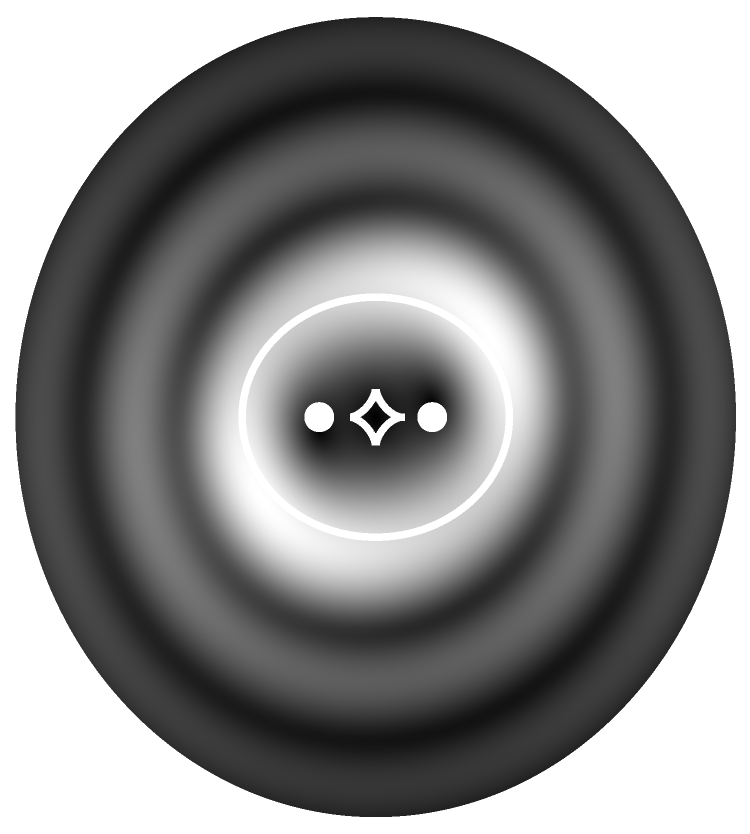}}
\centerline{\includegraphics[width=1.8cm,angle=-0]{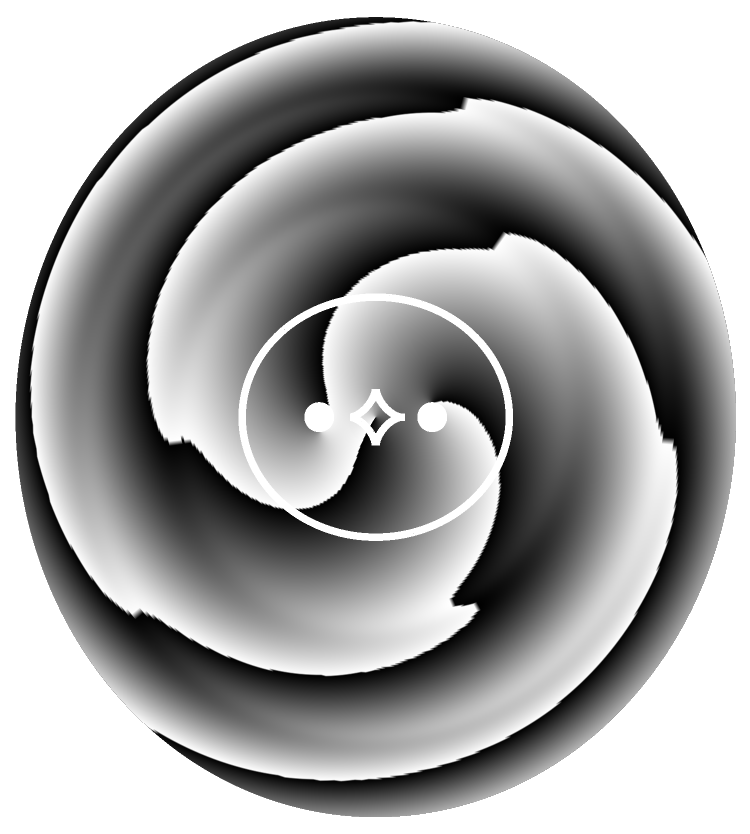}}
\end{minipage}
\hspace*{1mm}
\begin{minipage}[t]{1.8cm}
\centerline{\includegraphics[width=1.8cm,angle=-0]{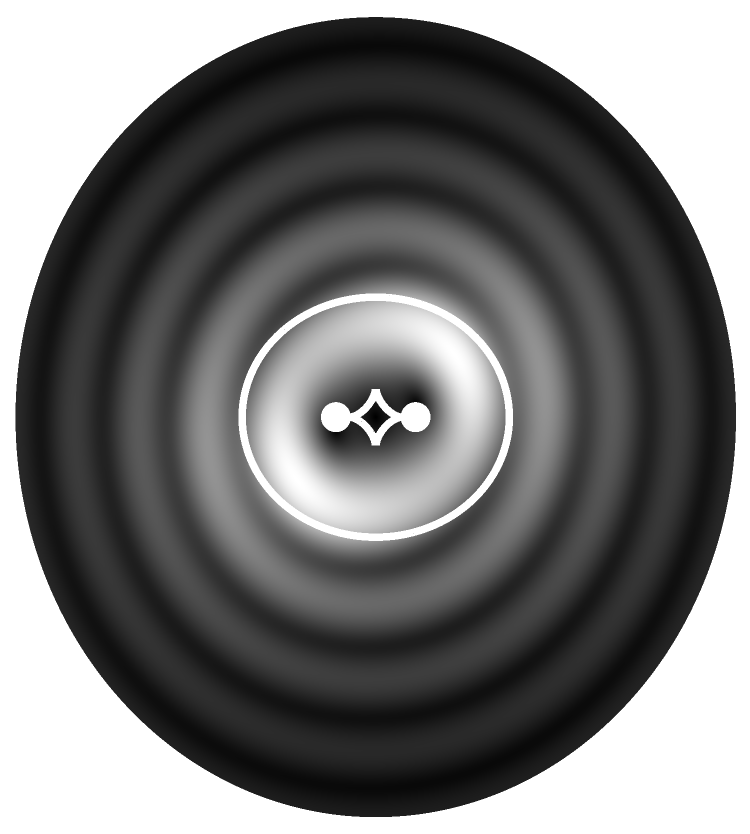}}
\centerline{\includegraphics[width=1.8cm,angle=-0]{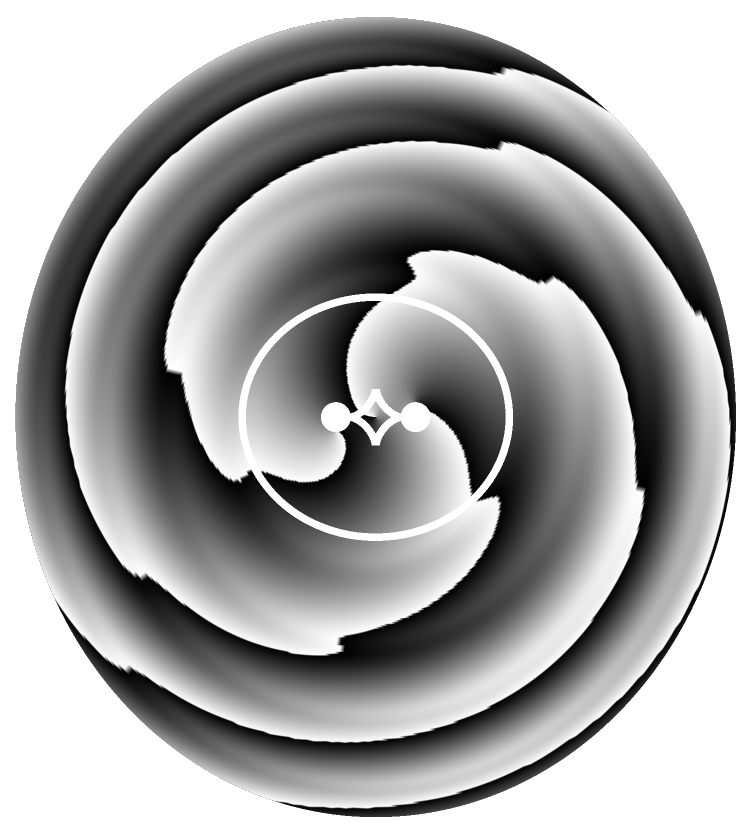}}
\end{minipage}
\caption{2D maps of the intensity (first row) and of the phase (second row)  distributions of the
wavefield produced by the diffraction of an ideal plane wave carrying out a topological charge $m=3$
by an elliptic aperture with $\chi=9/10$. Fresnel's numbers are $U=1$ (first column), $U=3$ (second column), 
$U=5$ (third column), and $U=7$ (fourth column). 
}
\label{Fig:Emme3VariesLowU}
\end{figure}

{ To give a single example of far-field evaluation, in Fig.~\ref{Fig:Emme3VariesLowU},  2D maps of the intensity (first row) 
and of the phase (second row) distributions of the wavefield produced by the diffraction of an ideal plane wave carrying out 
a topological charge $m=3$ by an elliptic aperture with $\chi=9/10$, are shown for ``low'' values of Fresnel's numbers,
i.e., for  $U <10$. 
}

\begin{figure}[!ht]
\centering
\begin{minipage}[t]{2.cm}
\centerline{\includegraphics[width=2.cm,angle=-0]{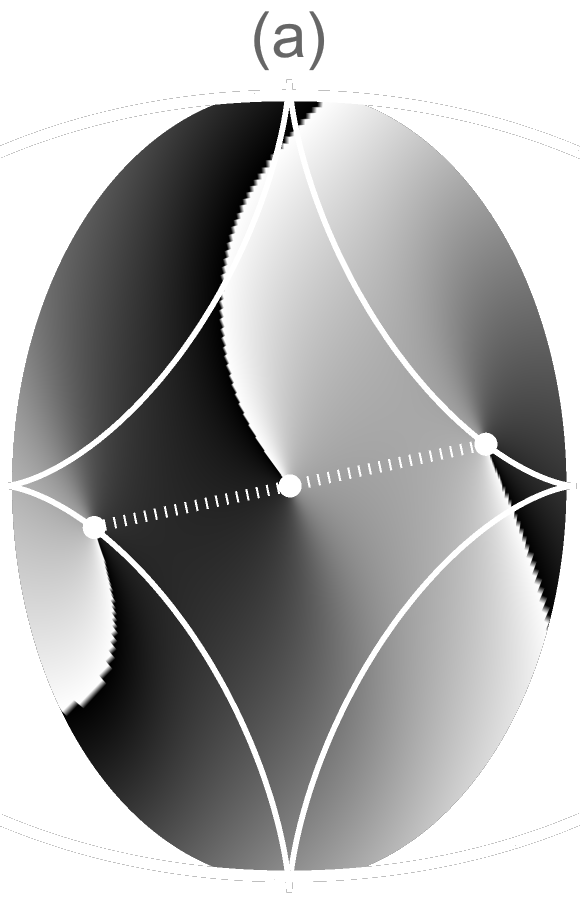}}
\end{minipage}
\hspace*{5mm}
\begin{minipage}[t]{2.5cm}
\centerline{\includegraphics[width=2.8cm,angle=-0]{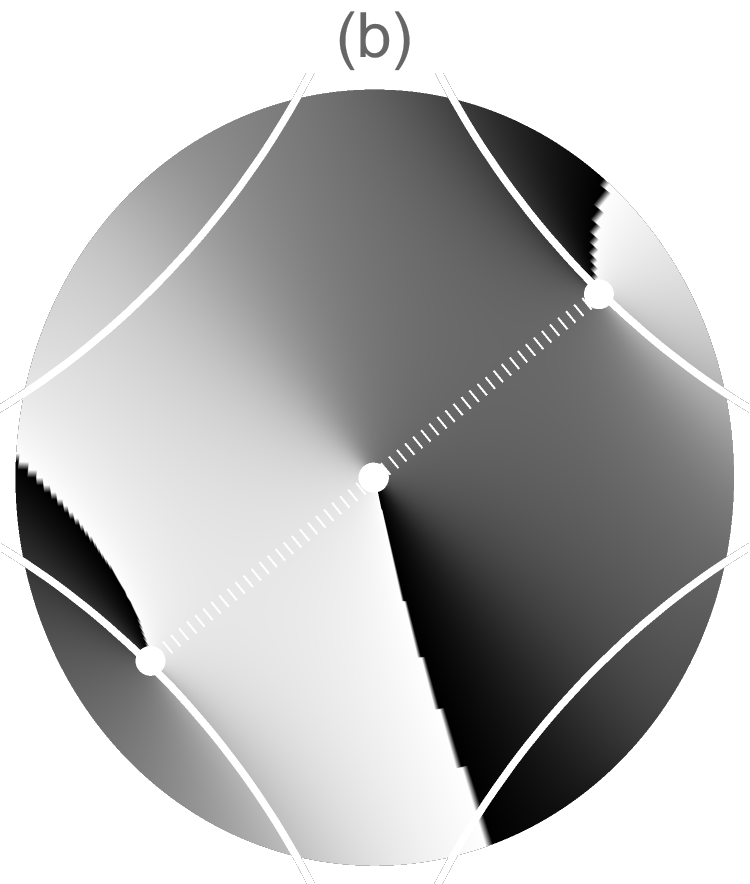}}
\end{minipage}
\caption{2D maps of the sole phase distributions 
for $\chi=7/10$ and $U=19/2$ (a) as well as for $\chi=925/1000$ and $U=89/2$ (b).
Now the white points denotes the { ``actual positions''}  of the three singularities. 
}
\label{Fig:Emme3U27}
\end{figure}

{ In all figures, the smooth white curve represents the elliptic aperture, while the  cusped white curve is the 
corresponding  evolute, defined through the parametric equation $\left\{(1-\chi^2) \cos^3t, \frac{1-\chi^2}\chi\sin^3t\right\}$, 
with~$t\in[0,2\pi]$. Moreover, the two white points correspond to the above defined ``far-field estimates'' of the outer singularities, where
in the present case $\bar p_x\,\simeq\,{2.074901}$.
From the figure, it can be appreciated how the far-field estimate of the outer singularities works reasonably 
well also for Fresnel's number values out of the far-field range, as for example $U=7$ certainly is. In other words, it is 
expected that the rectilinear trajectories penetrate inside the near-field zone, in agreement with some of the experimental 
results  provided in~\cite{Liu/Sun/Pu/Lu/2013} for a rhomb-shaped aperture.
On further increasing $U$, not only the outer singularities continue to get closer and closer, but it is also possible to appreciate a counterclockwise (for $m>0$) rotation, with respect to the $x$ axis,  
of the segment joining them. In other words, the outer singularity trajectories in the observation space $(\boldsymbol{r};U)$ deviate from the far-field rectilinear path, up to intersect the aperture evolute.}
{ To give an example, in Fig.~\ref{Fig:Emme3U27}, the phase distributions of the diffracted field are shown for $\chi=7/10$ and $U=19/2$ (a), as well as for 
$\chi=925/1000$ and $U=89/2$ (b). Note that, in this figure the white points now represent the ``actual positions'' of all three singularities. Moreover,
the Fresnel numbers have been chosen in order for the outer singularities to ``touch'' the astroid-shaped  caustics.
} 

\begin{figure}[!ht]
\centering
\begin{minipage}[t]{1.8cm}
\centerline{\includegraphics[width=1.8cm,angle=-0]{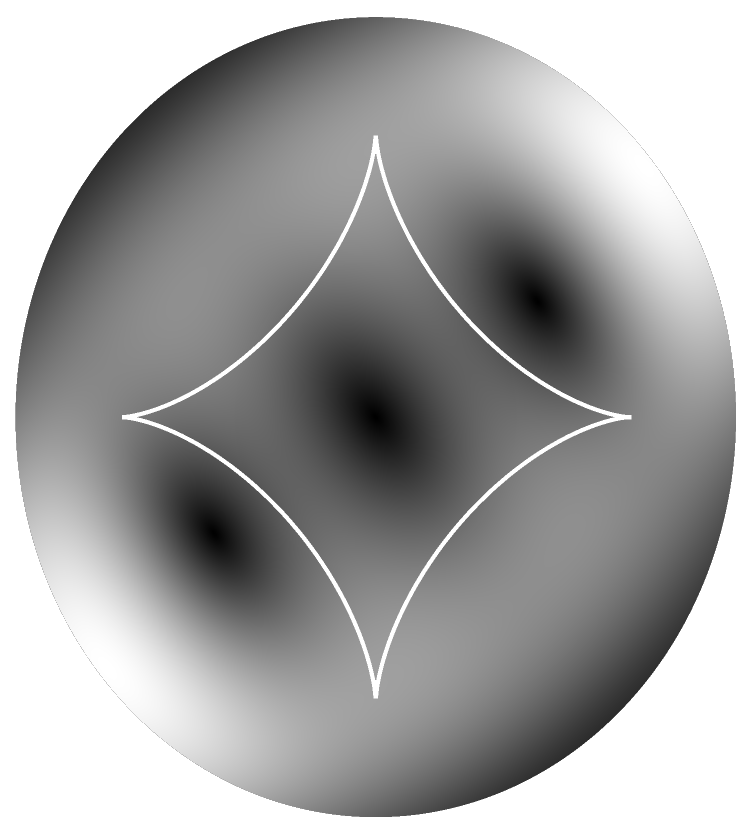}}
\centerline{\includegraphics[width=1.8cm,angle=-0]{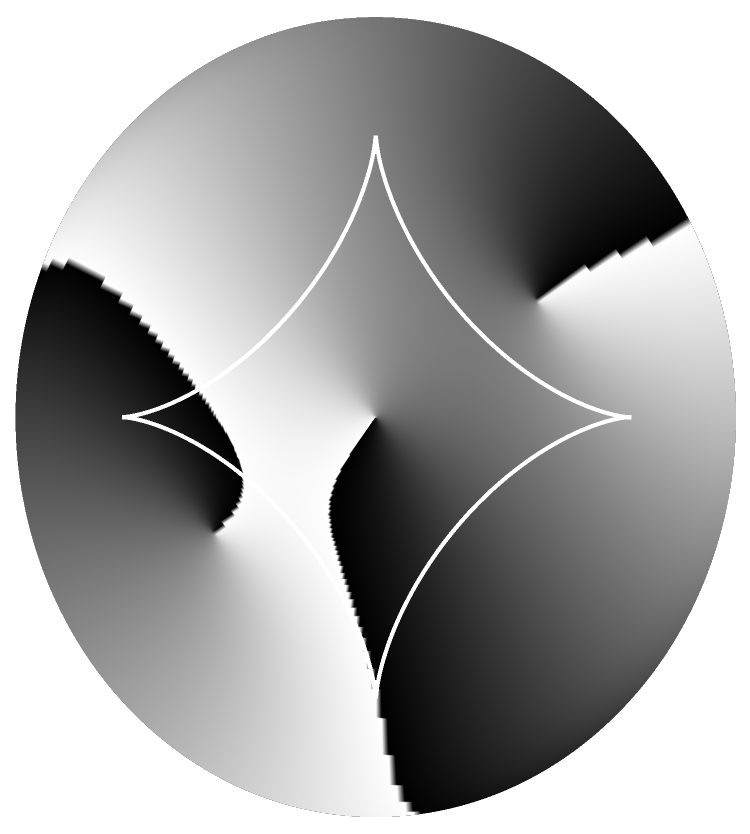}}
\end{minipage}
\hspace*{1mm}
\begin{minipage}[t]{1.8cm}
\centerline{\includegraphics[width=1.8cm,angle=-0]{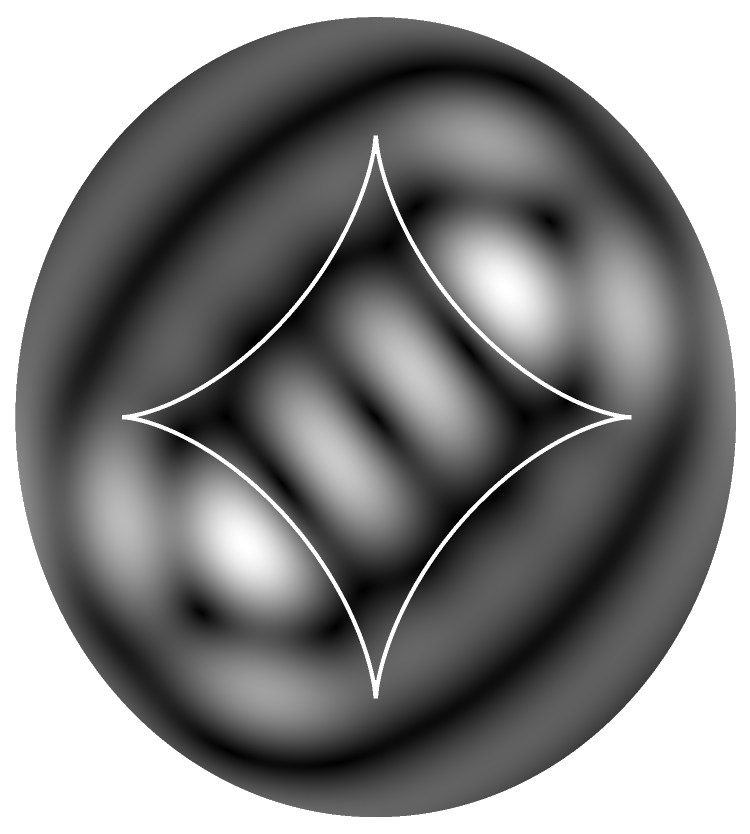}}
\centerline{\includegraphics[width=1.8cm,angle=-0]{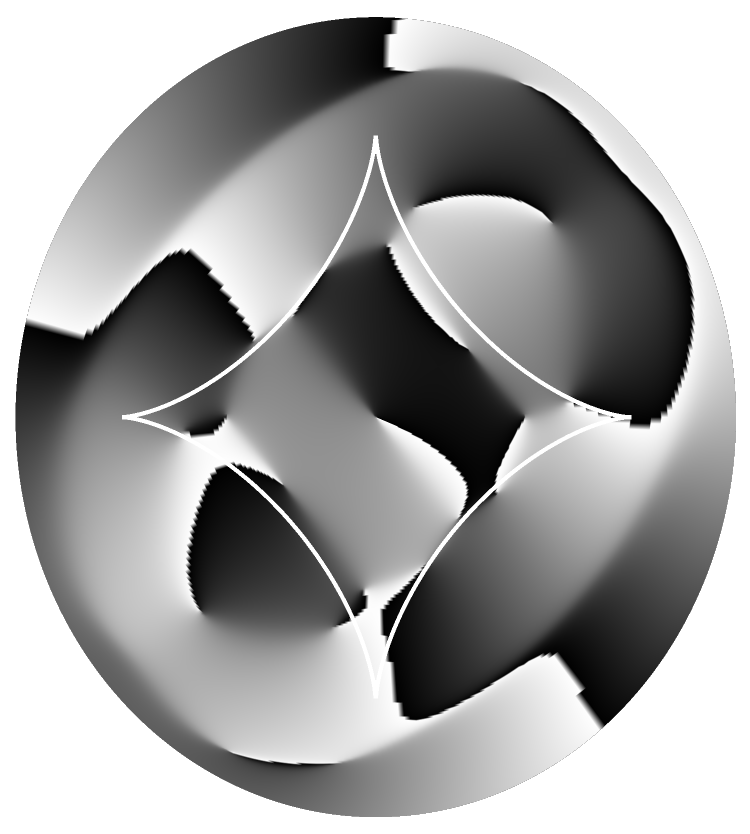}}
\end{minipage}
\hspace*{1mm}
\begin{minipage}[t]{1.8cm}
\centerline{\includegraphics[width=1.8cm,angle=-0]{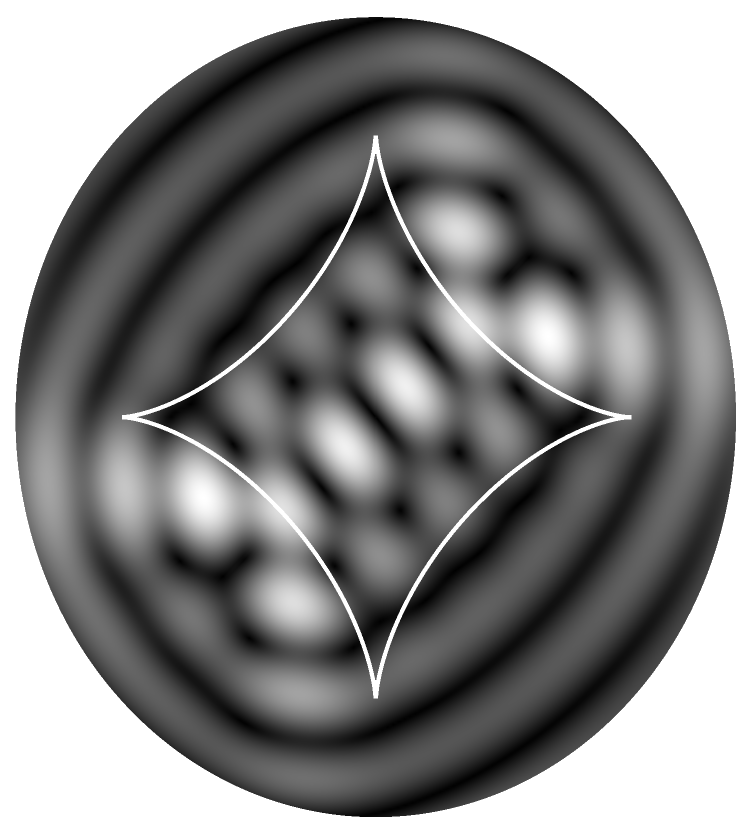}}
\centerline{\includegraphics[width=1.8cm,angle=-0]{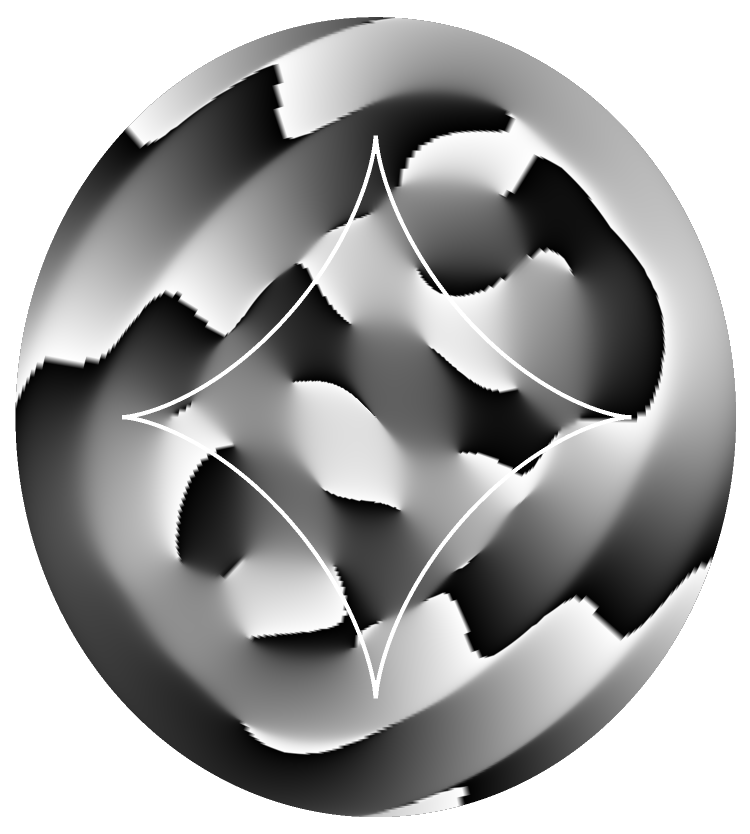}}
\end{minipage}
\hspace*{1mm}
\begin{minipage}[t]{1.8cm}
\centerline{\includegraphics[width=1.8cm,angle=-0]{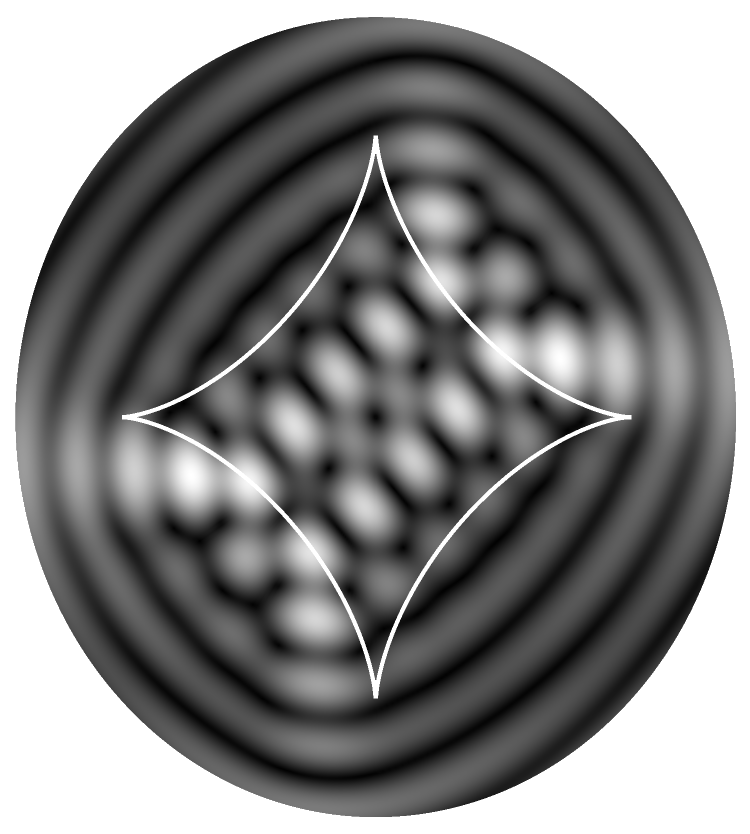}}
\centerline{\includegraphics[width=1.8cm,angle=-0]{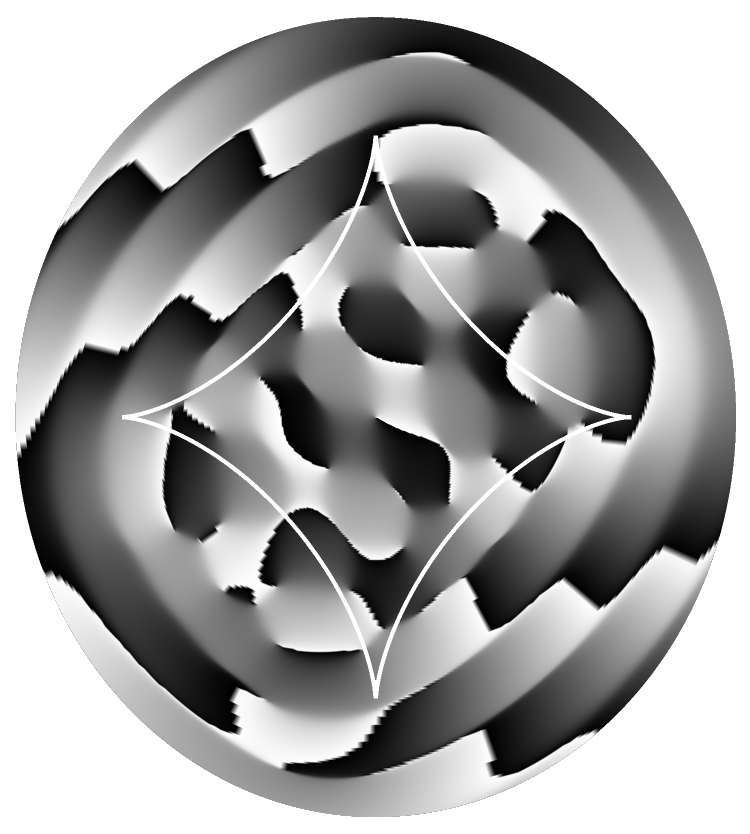}}
\end{minipage}
\caption{2D maps of the intensity (first row) and of the phase (second row)  distributions of the
wavefield produced by the diffraction of an ideal plane wave carrying out a topological charge $m=3$
by the elliptic aperture ($\chi=9/10$)  at different transverse 
observation planes. 
$U=19$ (first column), 
$U=41$ (second column), 
$U=59$ (third column), and
$U=79$ (fourth column).
}
\label{Fig:Emme3VariesU}
\end{figure}

On further increasing $U$,  all three singularities will be located inside the diamond-shaped region. 
To appreciate, at least qualitatively, such a new (topologically speaking) scenario, Fig.~\ref{Fig:Emme3VariesU} shows 2D maps of intensity (first row) and phase (second row) 
distributions  for $\chi=9/10$ and { for four increasing} different Fresnel's numbers.

\begin{figure}[!ht]
\centering
\begin{minipage}[t]{1.5cm}
\centerline{\includegraphics[width=1.5cm,angle=-0]{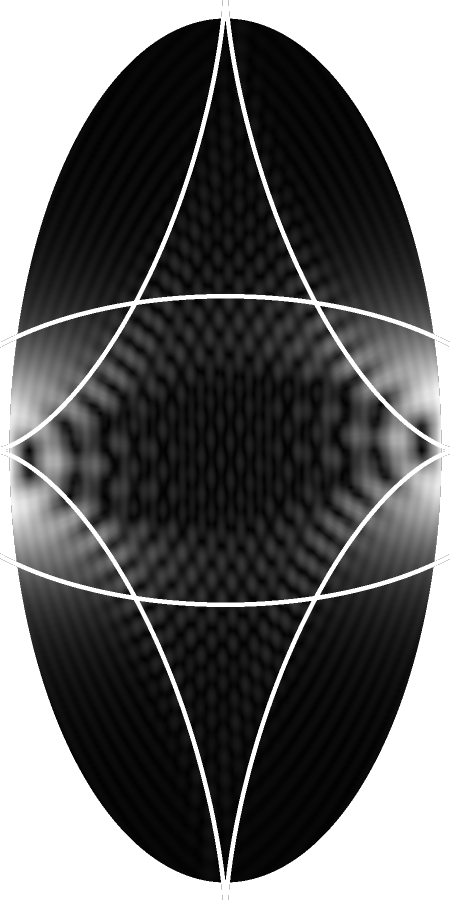}}
\centerline{\includegraphics[width=1.5cm,angle=-0]{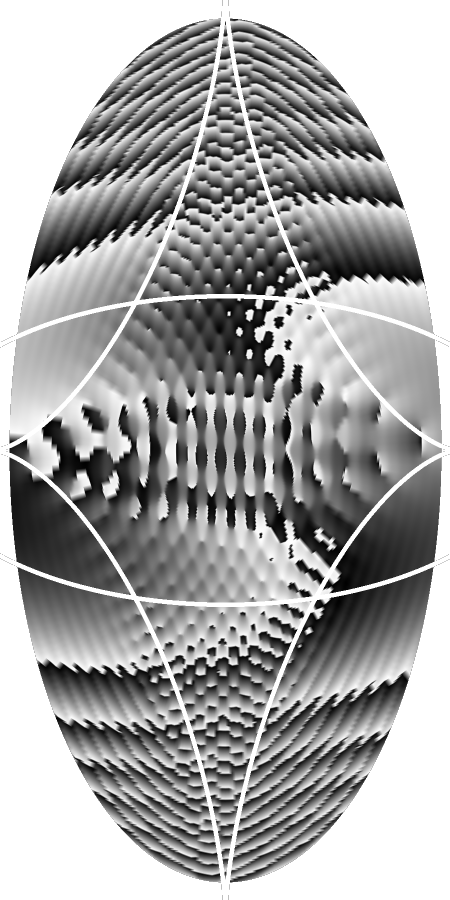}}
\end{minipage}
\hspace*{5mm}
\begin{minipage}[t]{2.cm}
\centerline{\includegraphics[width=2.0cm,angle=-0]{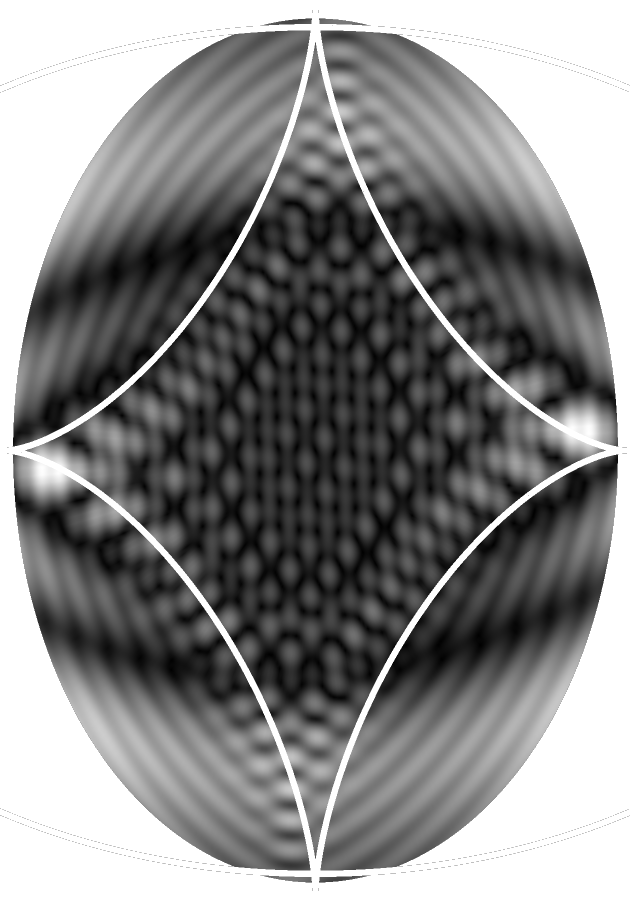}}
\centerline{\includegraphics[width=2.0cm,angle=-0]{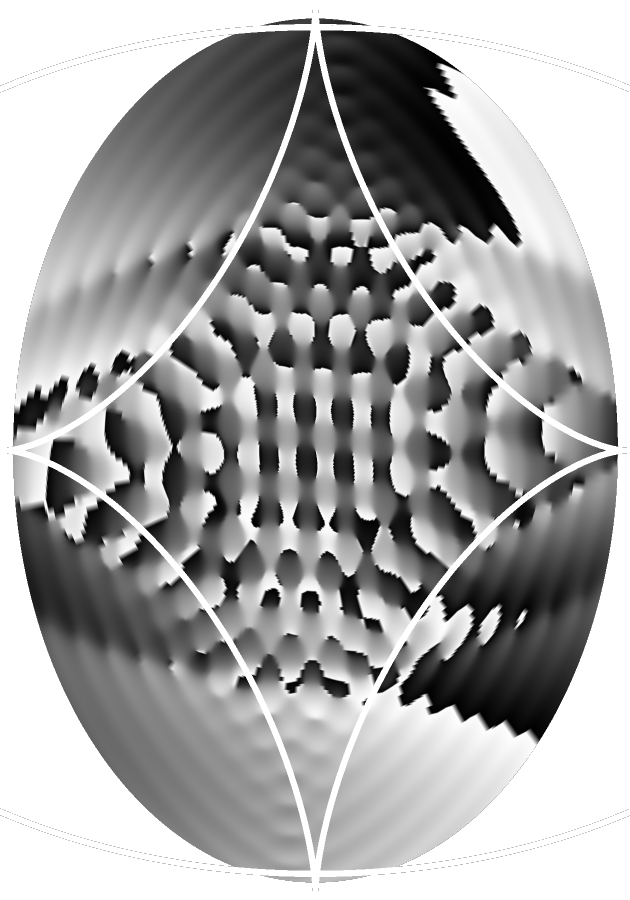}}
\end{minipage}
\hspace*{5mm}
\begin{minipage}[t]{2.4cm}
\centerline{\includegraphics[width=2.4cm,angle=-0]{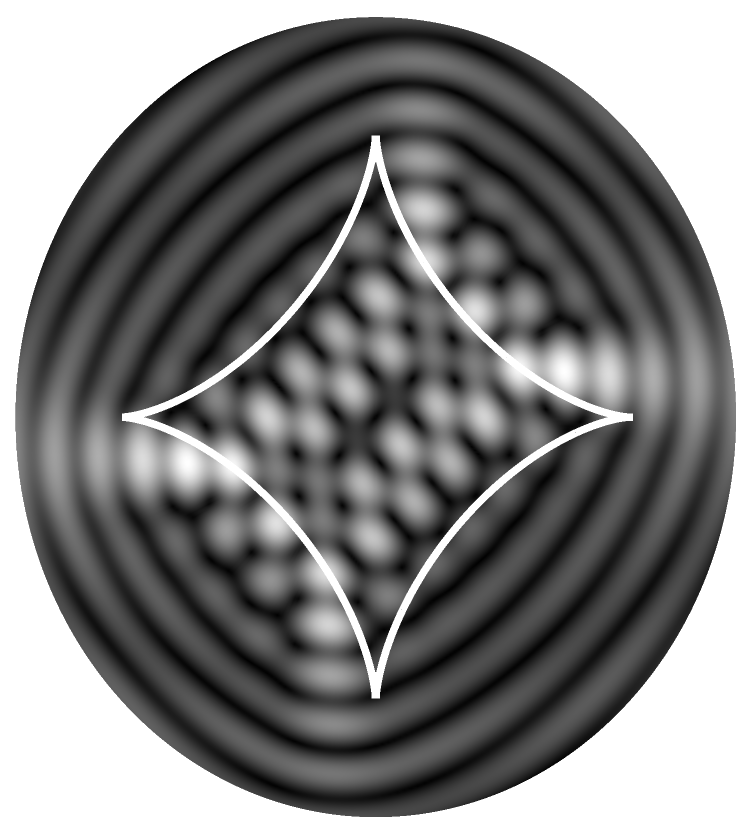}}
\centerline{\includegraphics[width=2.4cm,angle=-0]{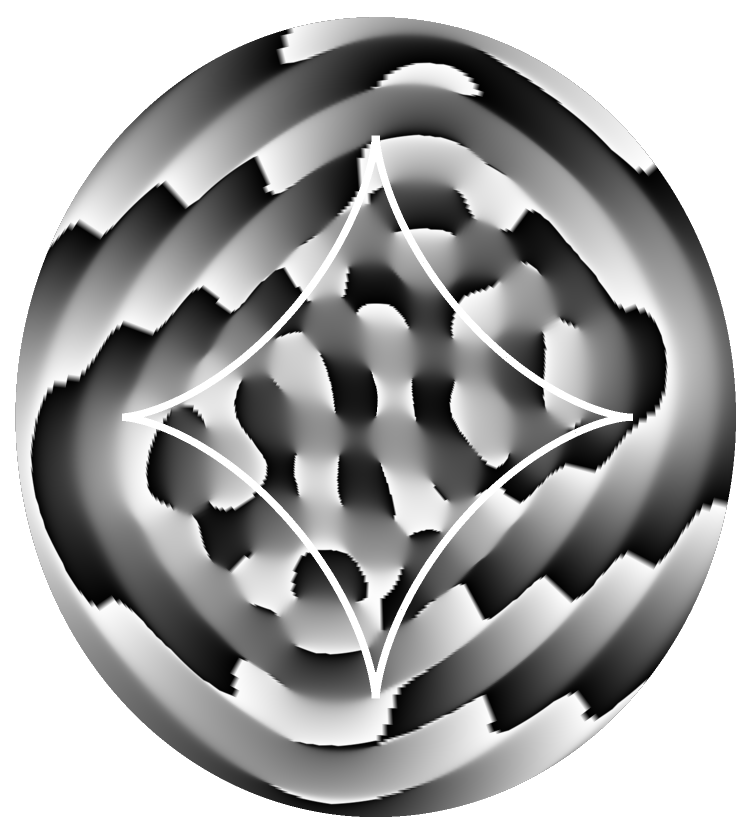}}
\end{minipage}
\caption{2D maps of the intensity (first row) and of the phase (second row)  distributions of the
wavefield produced by the diffraction of an ideal plane wave carrying out a topological charge $m=3$
at $U=101$,
by elliptic apertures  of different shapes.
$\chi=5/10$ (first column),
$\chi=7/10$ (second column),
and
$\chi=9/10$ (third column).
White cusped curves represent the corresponding aperture evolutes.}
\label{Fig:Emme3U131VariesChi}
\end{figure}

Several other numerical experiments, not shown here, have been carried out for different values of $\chi$ as well as of $m$.
All of them seem to confirm 
{ the key role the aperture evolute should play in discriminating the main topological features of the diffraction patterns.
A partial, qualitative confirm of such conjecture can be found  in~\cite{Hebri/Rasouli/Dezfouli/2019}, where the propagation of LG vortex beams apertured by smooth amplitude and phase Gaussian filters was studied.
In particular, the integration domain of the resulting diffraction integrals coincides with the whole $\mathbb{R}^2$ plane, 
while it is just the finiteness as well as the compactness of the integration domain (the elliptic aperture), that makes our intensity and phase patterns structurally different from those obtained 
in~\cite{Hebri/Rasouli/Dezfouli/2019}. 
Such a structural difference becomes more and more evident when Fresnel's number takes on large values.}
In Fig.~\ref{Fig:Emme3U131VariesChi} the 2D maps of intensity and phase are shown for $U=101$ and
for three different aperture shapes, namely $\chi=5/10$, $\chi=7/10$, and $\chi=9/10$. 
{ In  can be appreciated how the diffracted intensity rests onto the geometrical optics caustics (the cusped curve), which is ``dressed'' , to use a catastrophe optics terminology,
by the characteristic  Pearcey-based diffraction catastrophe~\cite{Borghi/2016}.}

\begin{figure}[!ht]
\centering
\begin{minipage}[t]{1.8cm}
\centerline{\includegraphics[width=1.8cm,angle=-0]{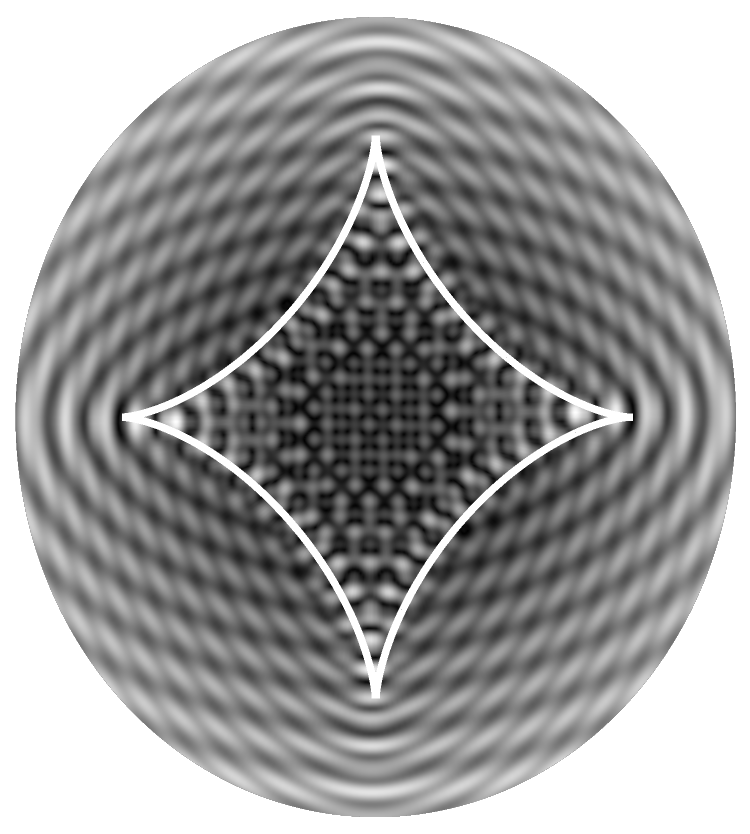}}
\centerline{\includegraphics[width=1.8cm,angle=-0]{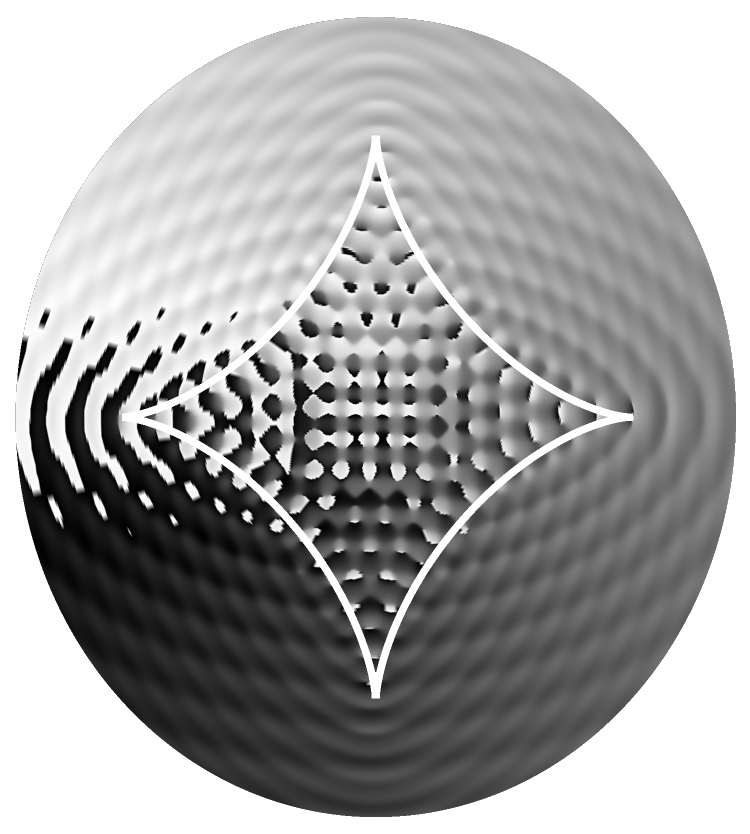}}
\end{minipage}
\hspace*{1mm}
\begin{minipage}[t]{1.8cm}
\centerline{\includegraphics[width=1.8cm,angle=-0]{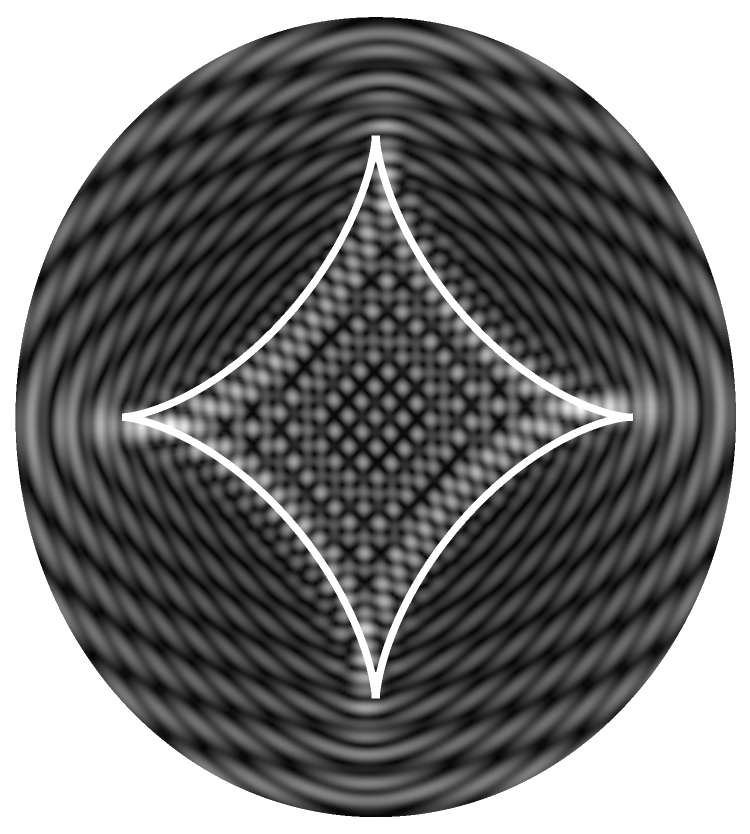}}
\centerline{\includegraphics[width=1.8cm,angle=-0]{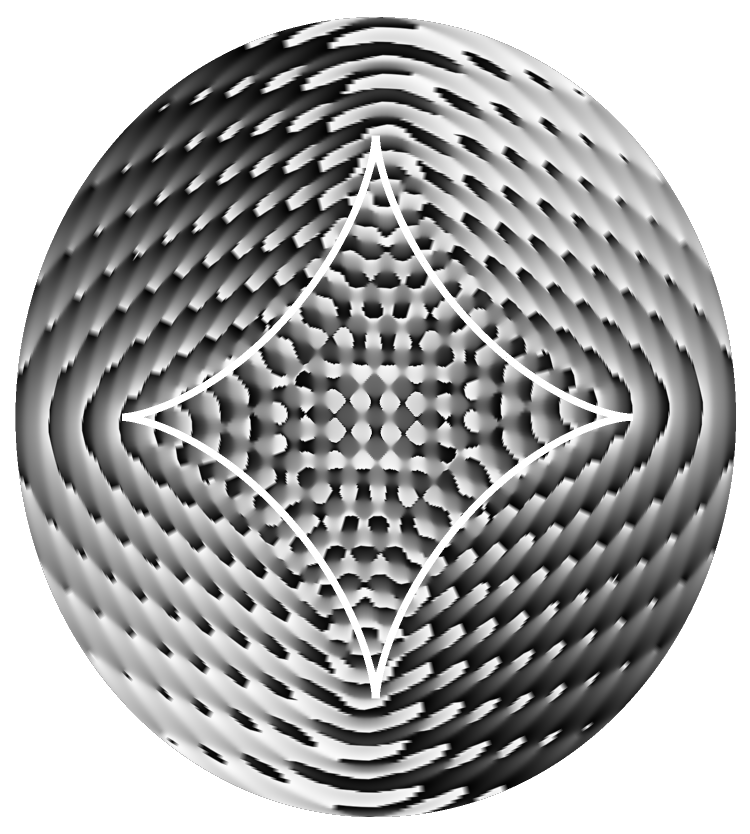}}
\end{minipage}
\hspace*{1mm}
\begin{minipage}[t]{1.8cm}
\centerline{\includegraphics[width=1.8cm,angle=-0]{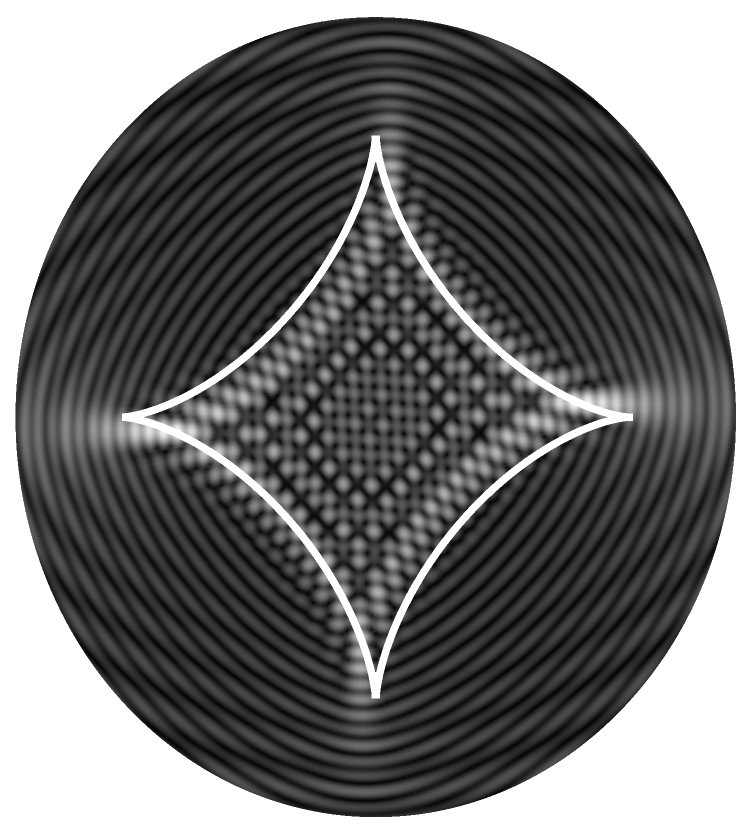}}
\centerline{\includegraphics[width=1.8cm,angle=-0]{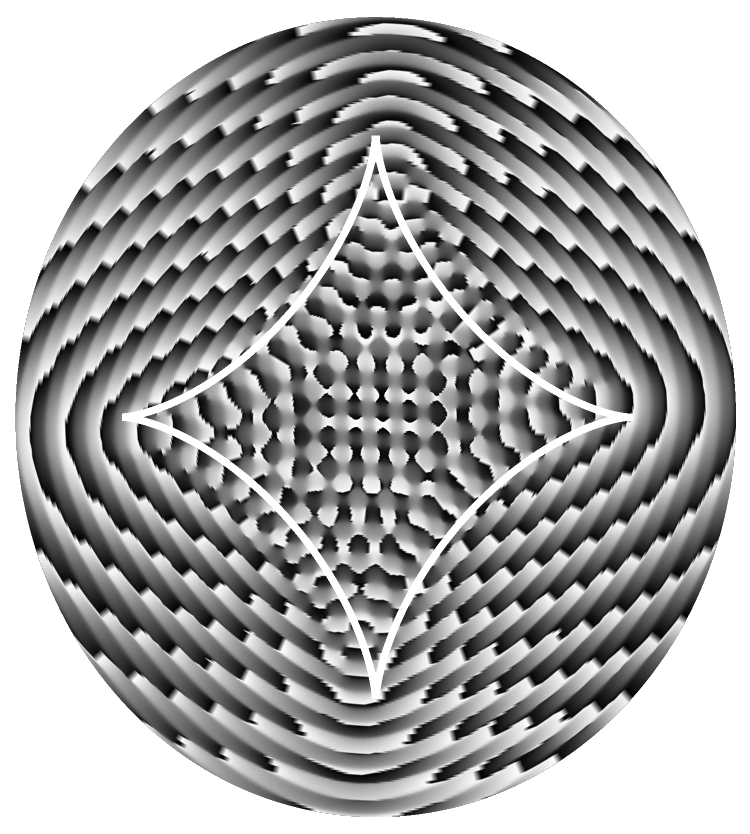}}
\end{minipage}
\hspace*{1mm}
\begin{minipage}[t]{1.8cm}
\centerline{\includegraphics[width=1.8cm,angle=-0]{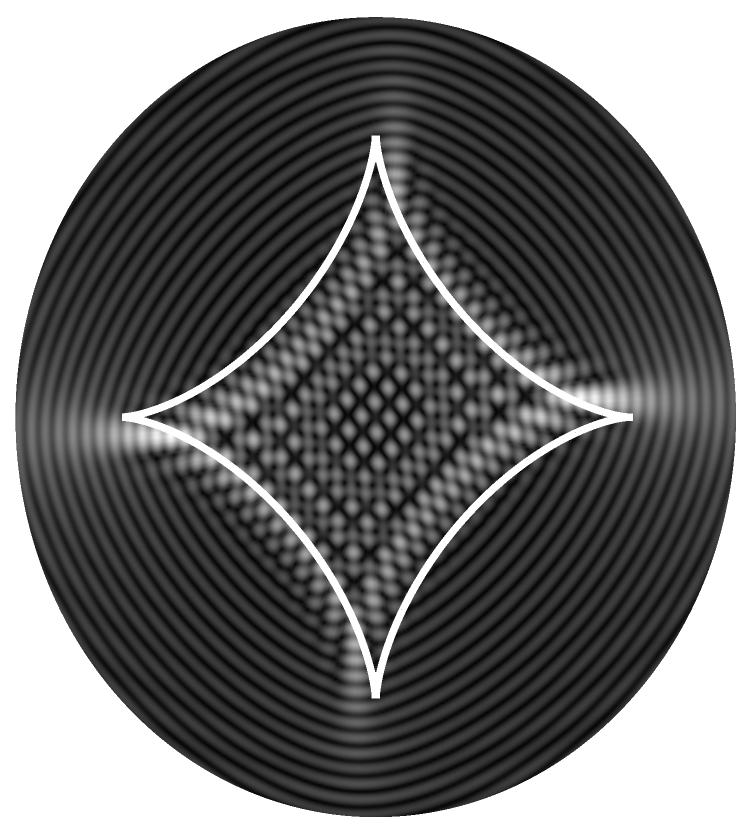}}
\centerline{\includegraphics[width=1.8cm,angle=-0]{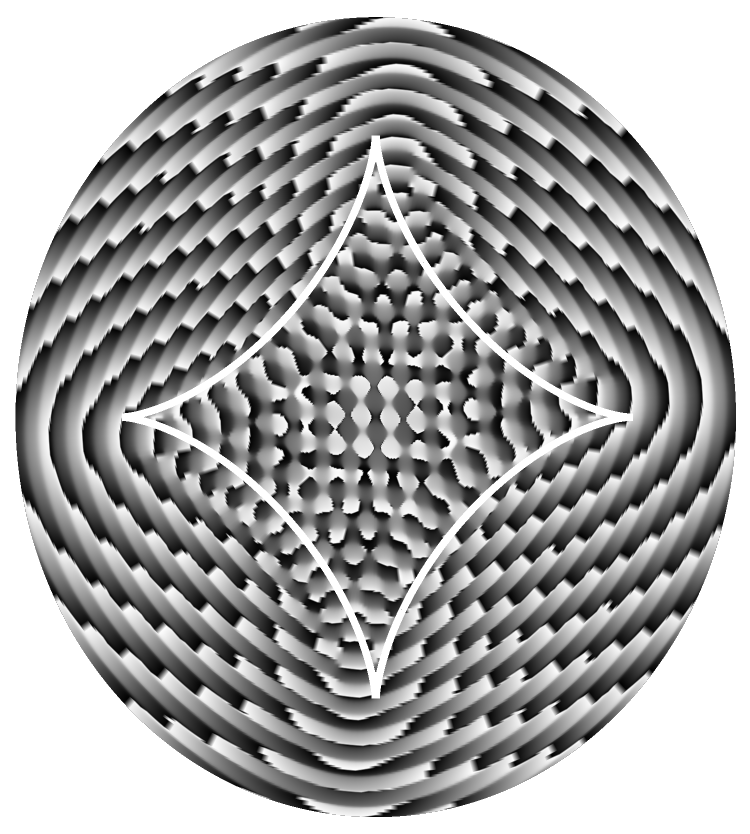}}
\end{minipage}
\caption{2D maps of the intensity (first row) and of the phase (second row)  distributions of the
wavefield produced by the diffraction of an ideal plane wave carrying out a topological charge $m=1$ (first column),
 $m=2$ (second column),  $m=3$ (third column), and  $m=4$ (fourth column), at $U=307$. The aperture shape is given by $\chi=9/10$.
White cusped curve represents the aperture evolute.}
\label{Fig:ChiNoveDecimiU211VariesEmme}
\end{figure}

{ 
Our last numerical experiment, showed in Fig.~\ref{Fig:ChiNoveDecimiU211VariesEmme}, is aimed at showing a visual comparison of diffraction patterns  generated 
by different topological charges, namely $m \in \{1, 2, 3, 4\}$. 
For simplicity, the chosen aperture corresponds to $\chi=9/10$, while the observation plane is located at $U=307$.}
{ It appears indubitably true that, when $U\gg 1$, the intensity and phase patterns inside the astroid display spectacularly regular lattices.
Thanks to the high resolution of the pictures (each of them contains $300\times 300$ points), it can be appreciated, for instance, a dependence of some features of the astroid-lattice topology 
on the {\em parity} of $m$ rather than its absolute value. For instance, due to symmetry reasons, odd values of $m$ must 
necessarily display one of the singularities exactly at the center of the diffraction pattern,  whereas even values of $m$ do not. This, in turn, implies that the corresponding intensity patterns
would share some topological features, as it can be checked simply by expanding the pictures.
Moreover, for the lowest value of $m=1$, the contribution coming from the central region of the illuminated aperture turns out to not negligible with respect to that coming from the aperture rim. 
On the contrary, for instance when $m=4$, the diffracted wavefield distribution could be ascribed to the sole edge waves, similarly to what happened in the plane-wave diffraction by opaque elliptic 
obstacles (see, for example, Fig.~10 of~\cite{Borghi/2014}).
}

In the present Letter, a semi-analytical algorithm for studying paraxial diffraction of vortex LG beams by centered elliptic apertures has  been developed.
{
The main conjecture of our paper is that the interaction between the singularities (phase vortices) carried out by the incident (plane wave) field and the singularities generated by the edge waves (the cusp-shaped evolute), 
is the main responsible for the resulting topological complexity of the diffraction patterns. 
In particular, our preliminary results seem to confirm the key role played by aperture's evolute in determining distinct topological scenarios, as far as intensity and phase distributions are concerned, 
 inside and outside of it, over a wide range of Fresnel's numbers.}

The nature of the present letter is merely computational. 
Our algorithm, based on rapidly convergent Fourier series, presents only one bottleneck: the numerical evaluation of the integrals into Eq.~(\ref{Eq:FresnelPropagatorConvolution.3,1.1}). 
However, it has been shown how high-resolution maps of the diffracted wavefield with, in principle, arbitrarily high accuracies can be (and have been) achieved even for Fresnel numbers
of the order of some hundreds. The availability of such an accurate and powerful computational tool should be welcomed by the optical community for several reasons. 
{ {\em In primis}, the results presented here, thanks to their high accuracy, could become a benchmark for testing new numerical methods able to deal with more general scenarios.
}

{ From an experimental perspective, we hope what is contained here could stimulate future activities. In particular, due to the intrinsic structural stability
of cusped singularities, we do hope experimental setups similar to other conceived in the past for similar purposes~\cite{Mourka/Baumgartl/Shanor/Dholakia/Wright/2011,Soifer/Kharitonov/Khonina/Volotovsky/2019,Xiao/Xie/Courvoisier/Hu/2022} could be conceived for experimentally exploring the vortex/caustics interaction here described. 
}

{ Theoretically speaking, several open problems would deserve deeper investigations, the most intriguing being the study of the complex topology of the diffraction patterns inside
the astroid. To this end, the joint use of our simulations  with more or less sophisticated algorithms of singularity tracking could greatly improve our understanding of the high-Fresnel number dynamics. 
That the edge waves generated at the aperture rim still focalize along the aperture evolute, seems to be valid also for the class of vortex beams here considered and not only under plane-wave illuminations~\cite{Borghi/2016}. This is something which could have not been taken for granted {\em a priori}. 
Accordingly,  our hope is that the beautiful interference patterns produced, in the limit of $U\gg 1$, inside the astroid could hopefully be decoded in the future by using the powerful language of 
catastrophe optics. To this end, a rigorous extension of paraxial BDW theory~\cite{Borghi/2016}  to plane waves carrying out optical vortices would certainly represent an important achievement to be reached.
}

\appendix

\section{Full Derivation of Eqs.~(4) and~(5) of the manuscript}
\label{Sec:IPart}

Consider a scalar disturbance orthogonally impinges onto a planar, opaque screen with a sharp-edge 
aperture $\CalA$, delimited by the closed boundary $\Gamma=\partial \CalA$ having the form of an ellipse.
The screen is placed at the plane $z=0$ of a suitable cylindrical reference frame $(\vettoreErre;z)$, whose $z$-axis coincides with the ellipse symmetry axis. 
Then, on denoting $\psi_0(\vettoreErre)$ the disturbance distribution at $z=0^-$, the field distribution, say $\psi$, at the observation point $P\equiv (\vettoreErre;z)$, 
with $z>0$ is given, within paraxial approximation and apart from an overall phase factor $\exp(\ii k z)$, by 
the following dimensionless version of Fresnel's integral:
\begin{equation}
\label{Eq:FresnelIntegral}
\begin{array}{l}
\displaystyle
\psi(\boldsymbol{r};U)\,=\,
-\frac{\mathrm{i}U}{2\pi}\,
\int_{\boldsymbol{\mathcal{A}}}\,
\mathrm{d}^2\,\rho\,
\psi_0(\boldsymbol{\rho})\,
\exp\left(\frac{\mathrm{i}U}{2}\,|\boldsymbol{r}-\boldsymbol{\rho}|^2\right)\,,
\end{array}
\end{equation}
where, in place of $z$, the Fresnel number $U=ka^2/z$ has been introduced, the symbol $a$ denoting  a characteristic length of the aperture size
which, in the present case, coincides with the ellipse major half-axis. Moreover, all transverse lengths, both across the aperture and the observation 
planes, will be normalized to $a$. In this way, the ellipse minor half-axis length turns out to be $\chi=\sqrt{1-\epsilon^2}$,  with $\epsilon \in [0,1]$ 
being ellipse's eccentricity.

Consider the following model of $\psi_0$:
\begin{equation}
\label{Eq:Gaussian.1}
\begin{array}{l}
\displaystyle
\psi_0(\boldsymbol{\rho})\,=\,\rho^m\,\exp(\ii\,m\,\phi)\,\exp\left(\ii \dfrac\gamma 2 \rho^2\right)\,, \quad\quad m \ge 0,\,\,\gamma \in \mathbb{C}\,,
\end{array}
\end{equation}
where $(\rho,\phi)$ denote polar coordinates of the transverse vector $\boldsymbol{\rho}$ across the aperture plane.
Here, the complex dimensionless parameter $\gamma$ gives account of the initial curvature (via 
$\mathrm{Re}\{\gamma\} \in \mathbb{R}$) as well as the transverse spot size (via 
$\mathrm{Im}\{\gamma\} \ge 0 $) of the impinging beam. 
The limit $\gamma\to 0$  represents the simpler, idealized scenario in which  a plane wave carrying out a 
topological charge $m$ is going to be diffracted by the sharp-edge elliptic aperture. 

Then, on substituting from Eq.~(\ref{Eq:Gaussian.1})  into Eq.~(\ref{Eq:FresnelIntegral}) we have
\begin{equation}
\label{Eq:FresnelPropagatorConvolution.1.1}
\begin{array}{l}
\displaystyle
\psi(\boldsymbol{r};U)\,=\,
-\frac{\mathrm{i}\,U}{2\pi}\,
\exp\left(\ii \dfrac U 2 r^2\right)\\
\\
\displaystyle\times
\int_{\boldsymbol{\rho}\in\mathcal{A}}\,
\mathrm{d}^2\rho\,\,
\rho^m\,\exp(\ii m\phi)\,
\exp\left(\ii \dfrac V 2 \rho^2\right)\,
\exp\left(-\mathrm{i}U\boldsymbol{r}\cdot\boldsymbol{\rho}\right)\,,
\end{array}
\end{equation}
where the complex parameter  $V=U+\gamma$, with $\mathrm{Re}\{\ii V\} \le 0$, has been introduced.


To evaluate the integral into~Eq.~(\ref{Eq:FresnelPropagatorConvolution.1.1}), 
Cartesian components of the transverse  vectors $\boldsymbol{\rho}$ and $\boldsymbol{r}$ will be represented as follows:
\begin{equation}
\label{Eq:EllipticHole.0.1}
\begin{array}{l}
\displaystyle
\boldsymbol{\rho} \,=\, (\xi\,\cos\alpha\,,\,\chi\,\xi\,\sin\alpha)\,,\qquad 0 \le \alpha \le 2\pi\,,\quad \xi \ge 0\,,
\end{array}
\end{equation}
and
%
%
\begin{equation}
\label{Eq:EllipticHole.1}
\begin{array}{l}
\displaystyle
\boldsymbol{r} \,=\, (\chi\,X\cos\beta\,,\,X\sin\beta)\,,\qquad 0 \le \beta \le 2\pi\,,\quad X \ge 0\,,
\end{array}
\end{equation}
respectively. It should be noted that there is an one-to-one correspondence between the pairs $(X,\beta)$ and $(r,\varphi)$, as well as 
between the pairs $(\xi,\alpha)$ and $(\rho,\phi)$. 
In particular, it is not difficult to prove that $r^2=X^2\,(1-\epsilon^2\,\cos^2\beta)$, $\boldsymbol{r}\cdot\boldsymbol{\rho}\,=\,\chi\xi X \cos(\alpha-\beta)$, and that
\begin{equation}
\label{Eq:EllipticHole.0.1.1}
\left\{
\begin{array}{l}
\displaystyle
\rho^2=\dfrac{1+\chi^2}2\xi^2+\dfrac{1-\chi^2}2\xi^2\,\cos 2\alpha\,,\\
\displaystyle\rho\exp(\ii\,\phi)\,=\,
\sum_{\sigma=\pm}\,
\eta_\sigma\,\exp(\sigma\ii\alpha)\,,
\end{array}
\right.
\end{equation}
where the quantities $\eta_\pm=(1\pm\chi)/2$ have been introduced.
%
%

Then, on substituting from Eqs.~(\ref{Eq:EllipticHole.0.1}) and~(\ref{Eq:EllipticHole.1}) into Eq.~(\ref{Eq:FresnelPropagatorConvolution.1.1}),
and on using Eqs.~(\ref{Eq:EllipticHole.0.1.1})  into account, straightforward algebra gives the following integral representation of the diffracted wavefield:
\begin{equation}
\label{Eq:FresnelPropagatorConvolution.2}
\begin{array}{l}
\displaystyle
\psi\,=\,
-\mathrm{i}U\chi\,
\exp\left(\frac{\mathrm{i}}{2}U\,r^2\right)
\int_0^1\,\mathrm{d}\xi\,\xi^{m+1}\,\exp\left(\ii V\dfrac{1+\chi^2}{4}\xi^2\right)\,\\
\\
\times\displaystyle
\frac 1{2\pi}\int_0^{2\pi}\mathrm{d}\alpha\,
\exp\left(\ii V\dfrac{1-\chi^2}{4}\xi^2\cos 2\alpha\right)\,\\
\times\displaystyle
\exp\left[-\ii UX\xi\chi\,\cos(\alpha\,-\,\beta)\right]\,
\left[\sum_{\sigma=\pm}\,\eta_\sigma\,\exp(\sigma\ii\alpha)\right]^m\,,
\end{array}
\end{equation}
where use has also been made of the fact that $\mathrm{d}^2\rho\,=\,\chi\,\xi\,\mathrm{d}\xi\,\mathrm{d}\alpha$.
It is now worth introducing the functions $\mathcal{J}_k({\rho},\varphi;\eta)$ 
defined as follows:
\begin{equation}
\label{Eq:FresnelPropagatorConvolution.2.1.1}
\begin{array}{l}
\displaystyle
\mathcal{J}_k({\rho},\varphi;\eta)\,=\,\\
\,=\,\displaystyle
\frac 1{2\pi}\int_0^{2\pi}\mathrm{d}\alpha\,
\exp(\ii k\alpha)
\exp\left({\mathrm{i}\eta}\cos 2\alpha\right)\,
\exp\left[-\mathrm{i}\,\rho\,\cos(\alpha\,-\,\varphi)\right]\,,
\end{array}
\end{equation}
with $k\in\mathbb{Z}$. A Fourier series representation of $\mathcal{J}_k$ can be given by using Jacobi-Anger's formula,  
\begin{equation}
\label{Eq:FourierApproach.1}
\begin{array}{l}
\displaystyle
\exp\left({\mathrm{i}\eta}\cos 2\alpha\right)\,=\,
\sum_{n=-\infty}^\infty\,\mathrm{i}^n\,J_n(\eta)\,\exp(\mathrm{i}2n\alpha)\,,
\end{array}
\end{equation}
which, once substituted into Eq.~(\ref{Eq:FresnelPropagatorConvolution.2.1.1}), after some algebra leads to
\begin{equation}
\label{Eq:FourierApproach.2}
\begin{array}{l}
\displaystyle
\mathcal{J}_k({\rho},\varphi;\eta)\,=\,
\ii^{-k}\exp(\ii k \varphi)\,
\sum_{n\in\mathbb{Z}}\,
\mathrm{i}^{-n}\,J_n(\eta)\,
J_{2n+k}(\rho)\,\exp(\mathrm{i}2n\varphi)\,,
\end{array}
\end{equation}
where use has been made of the well known integral representation of first kind Bessel functions,
\begin{equation}
\label{Eq:BesselJ}
\begin{array}{l}
\displaystyle
\frac 1{2\pi}\int_0^{2\pi}\mathrm{d}\alpha\,
\exp(\ii n\alpha)\,\exp\left(-\ii\,\rho\,\cos\alpha\right)\,=\,
\ii^{-n}\,J_{n}(\rho),\,\quad {{n\in \mathbb{Z}}\atop{\rho\in\mathbb{R}}}.
\end{array}
\end{equation}

Finally, on using Eqs.~(\ref{Eq:FresnelPropagatorConvolution.2}),~(\ref{Eq:FresnelPropagatorConvolution.2.1.1}), and~(\ref{Eq:FourierApproach.2}), long but in principle
straightforward algebra leads to the following representation of the diffracted wavefield $\psi$:
\begin{equation}
\label{Eq:FresnelPropagatorConvolution.3}
\begin{array}{l}
\displaystyle
\psi\,=\,
\displaystyle
-\dfrac{\mathrm{i}U\chi}2\,\exp\left(\frac{\mathrm{i}}{2}U\,r^2\right)\,\\
\times\displaystyle
\displaystyle\sum_{\ell=0}^m\,
\left({{m}\atop{\ell}}\right)\,\eta^\ell_+\eta^{m-l}_-\,
\Psi^m_{2\ell-m}\left[\dfrac{V(1+\chi^2)}{4},\,U\chi X,\,\dfrac{V(1-\chi^2)}4;\,\beta\right],
\end{array}
\end{equation}
where Newton's binomial theorem has been employed, the integration variable change $\xi^2\to \xi$ has been implemented,
and the  functions $\Psi^m_k$, defined as
\begin{equation}
\label{Eq:FresnelPropagatorConvolution.3,1}
\begin{array}{l}
\displaystyle
\Psi^m_k(a,b,c;\varphi)\,=\,
\int_0^1\,\mathrm{d}\xi\,\xi^{m/2}\,\exp(\ii a\xi)\,
\mathcal{J}_{k}\left(b\sqrt\xi,\varphi;c\,\xi\right)\,,\qquad\qquad -m \le k \le m\,,
\end{array}
\end{equation}
have been introduced.  In particular, on substituting from Eq.~(\ref{Eq:FourierApproach.2}) into Eq.~(\ref{Eq:FresnelPropagatorConvolution.3,1}), the latter
can be expressed through the following Fourier series:
\begin{equation}
\label{Eq:FresnelPropagatorConvolution.3,1.1}
\begin{array}{l}
\displaystyle
\Psi^m_k(a,b,c;\varphi)\,=\,
\ii^{-k}\exp(\ii k\varphi)\,\sum_{n\in\mathbb{Z}}\,
\ii^{-n} \exp(\ii 2n\varphi)\,\\
\\
\times
\displaystyle
\int_0^1\,\mathrm{d}\xi\,\xi^{m/2}\,\exp(\ii a\xi)\,
J_n(c\xi)\,J_{2n+k}(b\sqrt\xi)\,.
\end{array}
\end{equation}
%
Equations~(12)-(14) are equivalent to Eqs.~(3) and~(4) of the manuscript.

\section{Explicit analytical expression of the function defined by Eq.~(8) of the manuscript}
\label{Sec:IPart}

Consider the function defined by 
\begin{equation}
\label{Eq:FourierIntegral.1}
\begin{array}{l}
\displaystyle
\mathcal{F}_m(\boldsymbol{p})=\int_{\boldsymbol{\mathcal{A}}}\,
\mathrm{d}^2\,\rho\,
[\rho\,\exp(\mathrm{i} \phi)]^m
\exp\left(-\mathrm{i}\,\boldsymbol{p}\cdot\boldsymbol{\rho}\right)\,,\qquad m\in\mathbb{Z}\,.
\end{array}
\end{equation}

When $m=0$ (impinging plane wave), it is a well known result that 
\begin{equation}
\label{Eq:FourierIntegral.0}
\begin{array}{l}
\displaystyle
\mathcal{F}_0(\boldsymbol{p})\,=\,
\pi\chi\frac{2J_1(\sqrt{p^2_x+\chi^2p^2_y})}{\sqrt{p^2_x+\chi^2p^2_y}}\,,
\end{array}
\end{equation}
where the Cartesian representation of spatial frequency, i.e.,  $\boldsymbol{p}=(p_x,p_y)$, has been used.

For $m>0$, the FT evaluation becomes considerably harder,  due to the mismatch between the different symmetries 
related to $\psi_0$ (radial) and to the aperture (elliptical, of course). 

However, using partial derivativation under the Fourier integral, it is still possible to obtain an exact analytical expression
of $\mathcal{F}_m(\boldsymbol{p})$. To this end, the Cartesian representation  of $\boldsymbol{\rho}\equiv (\xi,\eta)$ is introduced, so that
\begin{equation}
\label{Eq:FourierIntegral.0.1}
\begin{array}{l}
\displaystyle
[\rho\,\exp(\mathrm{i} \phi)]^m\,=\,(\xi\,+\,\ii\,\eta)^m\,=\,
\sum_{k=0}^m\,\left({{m}\atop{k}}\right)\,\mathrm{i}^{m-k}\,\xi^k\,\eta^{m-k}\,,
\end{array}
\end{equation}
which, once substituted into Eq.~(\ref{Eq:FourierIntegral.1}), and after taking into account that
\begin{equation}
\label{Eq:FourierIntegral.0.2}
\begin{array}{l}
\displaystyle
\int\,\dd\xi\,\xi^k\,\exp(-\ii p_x\xi)\,=\,\ii^k\,
\dfrac{\partial^{k}}{\partial p_x^{k}}\,\int\,\dd\xi\,\exp(-\ii p_x\xi)
\end{array}
\end{equation}
and 
\begin{equation}
\label{Eq:FourierIntegral.0.3}
\begin{array}{l}
\displaystyle
\int\,\dd\eta\,\eta^{m-k}\,\exp(-\ii p_y\eta)\,=\,\ii^{m-k}\,
\dfrac{\partial^{m-k}}{\partial p_y^{m-k}}\,\int\,\dd\eta\,\exp(-\ii p_y\eta)\,,
\end{array}
\end{equation}
simple algebra eventually gives
\begin{equation}
\label{Eq:FourierIntegral.0.4}
\begin{array}{l}
\displaystyle
\mathcal{F}_m(\boldsymbol{p})\,=\,
(-1)^{m}\sum_{k=0}^m\,\left({{m}\atop{k}}\right)\,\mathrm{i}^{-k}\,
\dfrac{\partial^{m}}{\partial p_x^{k}\partial p_y^{m-k}}\,\mathcal{F}_0(\boldsymbol{p})\,.
\end{array}
\end{equation}

In all simulations of the manuscript,  the analytical evaluation of the function  $\mathcal{F}_m(\boldsymbol{p})$ has been done with the help of the latest release of {\em Mathematica} (release 13.3).
In particular, as far as $\mathcal{F}_3(\boldsymbol{p})$ is concerned, what we have found (with a nonnegligible effort) is the following closed-form formula:
\begin{equation}
\label{Eq:FourierIntegral.0.5}
\begin{array}{l}
\displaystyle
\mathcal{F}_3(\boldsymbol{p})\,=\,\pi\chi\,\dfrac{p_x+\ii \chi^2 p_y}{(p^2_x+\chi^2 p^2_y)^4}\,
\left\{
\mathcal{P}(p_x,p_y)\,J_0\left(\sqrt{p^2_x+\chi^2p^2_y}\right)\right.\\
\left.
\,-\,
\displaystyle
\mathcal{Q}(p_x,p_y)\,\dfrac{2J_1\left(\sqrt{p^2_x+\chi^2p^2_y}\right)}{\sqrt{p^2_x+\chi^2p^2_y}}
\right\}\,,
\end{array}
\end{equation}
where
\begin{equation}
\label{Eq:FourierIntegral.0.6}
\begin{array}{l}
\displaystyle
\mathcal{P}(p_x,p_y)\,=\,2\ii (p_x - \ii p_y \chi) (p_x +  \ii p_y \chi) \\
\times
(p_x^2 (-12 + p_x^2) + (-12 p_x^2 + 2 \ii p_x (-24 + p_x^2) p_y \\
+ (12 + p_x^2) p_y^2) \chi^2 +  p_y^2 (12 - p_x^2 + 2 \ii p_x p_y) \chi^4 - p_y^4 \chi^6)\,,
\end{array}
\end{equation}
and
\begin{equation}
\label{Eq:FourierIntegral.0.7}
\begin{array}{l}
\displaystyle
\mathcal{Q}(p_x,p_y)\,=\,(p^2_x+\chi^2p^2_y)\,\left[
\ii (p_x^2 (-24 + 5 p_x^2) \right. \\ 
 + (3 p_x^2 (-8 + p_x^2) + 16 \ii p_x (-6 + p_x^2) p_y  
 +  2 (12 + p_x^2) p_y^2) \chi^2 \\
\left. +  p_y^2 (24 - 2 p_x^2 + 16 \ii p_x p_y - 3 p_y^2) \chi^4 - 5 p_y^4 \chi^6)\right]\,.
\end{array}
\end{equation}

In particular, to determine the far-filed estimates of the singularity positions, it is sufficient to solve the trascendental equation 
$\mathcal{F}_3(p_x,0)=0$ which, after simple algebra leads to
\begin{equation}
\label{Eq:FourierIntegral.0.8}
\begin{array}{l}
\displaystyle
p_x (p_x^2 - 12 (1 + \chi^2)) J_0( p_x)\\
 + (24 - 5 p_x^2 - 3 (-8 + p_x^2) \chi^2) J_1( p_x)=0\,,
\end{array}
\end{equation}
whose least nonzero root $\bar p_x$ has been evaluated numerically for given values of $\chi$.

\section*{Acknowledgements}

{ I am indebted with the anonymous reviewers for their criticisms and suggestions, always aimed at 
improving the technical level of the paper.}
I wish to thank Turi Maria Spinozzi for his invaluable help during the preparation of the manuscript.

To Jari Turunen (1961-2023), in {\em Memoriam}.

\end{document}